# Iron Mobility during Diagenesis at Vera Rubin ridge, Gale Crater, Mars

J. L'Haridon[1], N. Mangold[1], A. A. Fraeman[2], J. R. Johnson[3], A. Cousin[4], W. Rapin[5], G. David[4], E. Dehouck[6], V. Sun[5], J. Frydenvang[7], O. Gasnault[4], P. Gasda[8], N. Lanza[8], O. Forni[4], P.-Y. Meslin[4], S. P. Schwenzer[9], J. Bridges[10], B. Horgan[11], C. H. House[12], M. Salvatore[13], S. Maurice[4], R. C. Wiens[7]

[1] *Laboratoire de Planétologie et Géodynamique, UMR6112, CNRS, Univ Nantes, Univ Angers, Nantes, France*
[2] *Jet Propulsion Laboratory, Pasadena, California, USA*
[3] *Johns Hopkins University Applied Physics Laboratory, Laurel, Maryland, USA*
[4] *IRAP, UPS, OMP, Toulouse, France*
[5] *California Institute of Technology, Pasadena, California, USA*
[6] *Univ Lyon, Univ Lyon 1, ENSL, CNRS, LGL-TPE, F-69622, Villeurbanne, France*
[7] *University of Copenhagen, Copenhagen, Denmark*
[8] *Los Alamos National Laboratory, Los Alamos, New Mexico, USA*
[9] *Open University, Milton Keynes, UK*
[10] *University of Leicester, UK*
[11] *Purdue University, USA*
[12] *Dept of Geosciences, Pennsylvania State University, USA*
[13] *North Arizona University, Flagstaff, USA*

**HIGHLIGHTS**

→ Images from the *Curiosity* rover show the presence of dark-toned diagenetic features at Vera Rubin ridge

→ ChemCam analyses of these features point towards a Fe-oxide composition, consistent with crystalline hematite

→ Depletion of Fe and Mn in bleached halos surrounding the high-Fe diagenetic features indicates mobility of Fe and Mn during the later stages of diagenesis

**ABSTRACT**

The *Curiosity* rover investigated a topographic structure known as Vera Rubin ridge, associated with a hematite signature in orbital spectra. There, *Curiosity* encountered mudstones interpreted as lacustrine deposits, in continuity with the 300 m-thick underlying sedimentary rocks of the Murray formation at the base of Mount Sharp. While the presence of hematite ($\alpha$-Fe2O3) was confirmed in-situ by both Mastcam and ChemCam spectral observations and by the CheMin instrument, neither ChemCam nor APXS observed any significant increase in $FeO_T$ (total iron oxide) abundances compared to the Murray formation. Instead, *Curiosity* discovered dark-toned diagenetic features displaying anomalously high $FeO_T$ abundances, commonly observed in association with light-toned Ca-sulfate veins but also as crystal pseudomorphs in the host rock. These iron-rich diagenetic features are predominantly observed in "grey" outcrops on the upper part of the ridge, which lack the telltale ferric signature of other Vera Rubin ridge outcrops. Their composition is consistent with anhydrous Fe-oxide, as the enrichment in iron is not associated with enrichment in any other elements, nor with detections of volatiles. The lack of ferric absorption features in the ChemCam reflectance spectra and the hexagonal crystalline structure associated with dark-toned crystals points toward coarse "grey" hematite. In addition, the host rock adjacent to these features appears bleached and show low-$FeO_T$ content as well as depletion in Mn, indicating mobilization of these redox-sensitive elements during diagenesis. Thus, groundwater fluid circulations could account for the remobilization of iron and recrystallization as crystalline hematite during diagenesis as well as color variations observed in the Vera Rubin ridge outcrops.

**PLAIN SUMMARY**

The NASA rover Curiosity investigated Vera Rubin ridge, a specific landform within the Gale crater on Mars. Scientific missions in orbit around the planet had previously discovered high concentrations of hematite on top of the ridge, an iron-oxide mineral that commonly forms in water. However, it was not clear from orbit if such conditions existed at the time of the deposition of the sediments (around 3.5 billion years ago) or occurred much later during "diagenesis", after deposition of the sediments and up to their transformation into rocks. On the surface, the rover did not observe significant differences between the ridge and the terrains encountered before it, except for small, dark geologic features that formed during diagenesis. Their analysis by the ChemCam instrument revealed that these features are composed of hematite—the same iron-oxide mineral that was observed from orbit—and interestingly, that the iron required to form them was removed from the adjacent rocks by groundwaters. Thus, it appears that groundwaters played an important role in shaping Vera Rubin ridge, and thus partially obscure interpretations on the environmental conditions that existed on the surface of Mars at the time of sedimentation.

# 1. Introduction

In 2012, the Mars Science Laboratory (MSL) *Curiosity* rover landed on the plains of a ~155 km diameter impact crater, Gale crater, north of the 5 km high sedimentary mound (Aeolis Mons, informally referred to as Mount Sharp) that sits at its center (Grotzinger *et al.*, 2012). After 1800 sols (sol = Martian day) and ~17 km of traverse across regolith and outcrops of fluvio-lacustrine deposits (Grotzinger *et al.*, 2014, 2015; Vasavada *et al.*, 2014) and of local remnant of a draping eolian unit (Banham *et al.*, 2018), the rover reached Vera Rubin ridge (VRR), on the lower slopes of Mount Sharp (Figure 1). This ~6.5 km long and ~200 m wide topographic sedimentary structure trends northeast-southwest on the fringe of the phyllosilicate-bearing trough (Glen Torridon) and overlying sulfate deposits (Anderson, 2010; Milliken, Grotzinger and Thomson, 2010; Fraeman *et al.*, 2016). It also displays higher thermal inertia and greater resistance to erosion compared to the adjacent terrains, from both orbital and in-situ perspectives.

Hyperspectral observations from the Compact Reconnaissance Imaging Spectrometer for Mars (CRISM) instrument onboard Mars Reconnaissance Orbiter (MRO) revealed a distinctive crystalline hematite signature on the upper-most part of the ridge (Fraeman *et al.*, 2013). The hematite-rich nature of the ridge was interpreted as evidence for the involvement of redox-driven processes (iron oxidation) in its formation, either related to primary deposition (e.g. redox interface in the lacustrine setting or soil formation) or post-depositional processes during diagenesis (e.g. oxidation by neutral to acidic groundwater) (Fraeman *et al.*, 2013). Vera Rubin ridge thus provides a unique opportunity to investigate redox processes on Mars, but also to compare orbital observations and interpretations against the ground truth. As such, this topographic structure constitutes a key milestone in the rover's investigation of the Gale crater's geological history and its implications for past habitability on the planet (Fraeman et al., this issue).

The VRR rocks exhibit a range of geological features that formed during diagenesis, i.e. after deposition of the sediments and through their burial and induration into sedimentary rocks. The pervasive light-toned Ca-sulfate veins encountered in all geological units crossed by the rover traverse so far (Nachon *et al.*, 2014; Rapin *et al.*, 2016; L'Haridon *et al.*, 2018) are also abundant throughout Vera Rubin ridge (Figure 1). Crystal molds, nodules, and concretions are observed in close association with Ca-sulfate-filled veins. In this study, we report results from the ChemCam instrument with respect to small-scale diagenetic features observed at Vera Rubin ridge. Lastly, we discuss the potential implications of these findings with respect to previous observations in Gale crater and argue for the significant role played by diagenetic processes in the formation of the Vera Rubin ridge sedimentary rocks.

# 2. Methods

## 2.1 ChemCam instrument

ChemCam (Maurice *et al.*, 2012; Wiens *et al.*, 2012) is composed of a Laser-Induced Breakdown Spectroscopy (LIBS) system coupled with a Remote Micro-Imager (RMI; (Le Mouélic *et al.*, 2015)). The LIBS technique consists in focusing a pulsed laser on the target material (located up to 7 m from the mast of the rover), which leads to the formation of a plasma. The light emitted from the decay of its excited atoms, ions, and molecules to lower electronic states is then analyzed by three spectrometers (covering the spectral ranges of 240-342 nm, 382-469 nm, and 474-906 nm) that resolve emission peaks used for elemental analysis. ChemCam analyses consist of several closely spaced observation points, each typically composed of 30 laser shots on a given point. In this process, dust is removed from the surface of the target by the first few laser shots, thus providing the composition of the dust-free surface target material.

ChemCam provides fine-scale chemical analyses especially suited to characterizing isolated geological features such as concretions, nodules, or fracture-fills. Indeed, the ablated area covers about 300 to 600 µm in diameter, and up to a few micrometers per laser shot in depth depending on the targeted material (Maurice *et al.*, 2012), complementary to the bulk analyses of the other instruments on the rover. While ChemCam was not designed to provide information on mineralogy, mineral phases can still be inferred from ChemCam chemical compositions for geological features when their size is comparable to or larger than the laser spot size. Even at these scales, ChemCam may analyze a mixture of compositions if the LIBS point is located at the interface between distinct features. For instance, points on light-toned veins commonly show mixing between Ca-sulfate from the veins and the surrounding host rock composition (Nachon *et al.*, 2014, 2016; Rapin *et al.*, 2016; L'Haridon *et al.*, 2018).

Passive reflectance spectra (without active laser use) are acquired after each ChemCam observation, recording the Martian surface radiance in the visible/near-infrared (400–840 nm) using ChemCam spectrometers. The passive reflectance data can guide interpretations on mineralogy as it provides additional information on the presence of $Fe^{3+}$-bearing mineral phases and crystallinity, such as hematite (Johnson *et al.*, 2015, 2016), which is otherwise inaccessible from LIBS analyses. The angular field of view of ~0.65 mrad of the spectrometer corresponds to a 2 mm diameter area on the surface at 3 m distance from the rover (50% of ChemCam targets are within 2 to 3 m from the rover and 45% within 3-5 m (Maurice *et al.*, 2016)). It is thus wider than the LIBS ablation area.

## 2.2 Elemental composition and quantification

ChemCam is able to detect all major elements; their abundances are quantified in weight percent of oxides ($SiO_2$, $TiO_2$, $Al_2O_3$, total iron oxide expressed as $FeO_T$, $MgO$, $CaO$, $Na_2O$ and $K_2O$) using a combination of two multivariate analytical methods – namely the "Partial Least Squares" (PLS) and "Independent Component Analysis" (ICA) methods – that results in the Multivariate Oxide Composition (MOC) dataset (Clegg *et al.*, 2017). Expansion of the ChemCam calibration database from 66 to 408 standards and associated recalibration was completed in 2015 (Wiens and Maurice, 2015; Clegg *et al.*, 2017) thus refining the quantification of major elements with the ChemCam instrument. Accuracy is estimated through the calculation of the root mean squared error of prediction (RMSEP) for a representative test set of standards for each major element. Since accuracy varies with elemental abundances, a sub-model approach has been developed in order to provide more accurate abundances across the distinct compositional ranges observed for each element on Mars (Anderson *et al.*, 2017; Clegg *et al.*, 2017). The measurement precision is mainly defined by ChemCam's repeatability using point-to-point and the shot-to-shot prediction variations either on the target itself, selected homogeneous targets or the calibration targets (Clegg *et al.*, 2017, and references therein).

A number of minor and trace elements detected by ChemCam are also quantified using a univariate method, in which a single emission peak is calibrated for the abundance of that element. These include Mn (Lanza *et al.*, 2014), Sr, Li, Ba, Rb (V. Payré *et al.*, 2017), Cu (V Payré *et al.*, 2017), Zn (Lasue *et al.*, 2016), B (Gasda *et al.*, 2017), and F (Forni *et al.*, 2015). In addition, the hydration state for specific compositions can be derived from the H signal using recent calibration efforts (Rapin *et al.*, 2016). Other volatile elements such as S, P, and Cl are so far only observed qualitatively, but are often used for chemical and mineralogical diagnostic studies (Nachon *et al.*, 2016; Meslin *et al.*, 2018), although work is currently in progress to quantify them systematically (Meslin *et al.*, 2016; Anderson *et al.*, 2017; Clegg *et al.*, 2018; Thomas *et al.*, 2018; Rapin *et al.*, 2019).

High-Fe compositions are, however, still poorly represented in the current database (Clegg *et al.*, 2017), with only 2 samples with > 40 wt.% $FeO_T$, which notably leads to an underestimation of $FeO_T$ in Fe-oxide dominated compositions on Mars. Recent laboratory LIBS analyses of hematite-basalt mixtures and iron meteorite samples yielded improved predictions for $FeO_T$ (David et al., this issue). While this laboratory work was not yet incorporated in the ChemCam quantification models at the time of writing, the improved calibration for $FeO_T$ will be considered here to characterize the variability of $FeO_T$ associated with diagenesis on VRR, especially for high-$FeO_T$ abundances. Also, note that the abundances in $FeO_T$ reported in the paper corresponds to the total iron oxide content, as ChemCam is unable to distinguish native, ferrous and ferric iron. This study includes the ChemCam data acquired from sol 1809 to sol 2300 (431 ChemCam targets and 3554 individual points), covering the whole exploration of VRR. ChemCam observations on float rocks, meteorites, wind-blown sand, drill tailings, and drill material dump piles were not included in the data set of this study as they do not contribute to the purpose of this study.

## 2.3   Mastcam and MAHLI cameras

The VRR outcrops were documented by the Mast Cameras (Mastcam) and Mars Hand Lens Imager (MAHLI). The two Mastcam cameras (34 mm and 100 mm focal lengths), located on the mast, provide contextual color images for ChemCam observations at the outcrop scale, with a 150 µm/pixel resolution at 2 m (Malin *et al.*, 2017). MAHLI, mounted on the robotic arm of the rover, provides high-resolution (up to 14 µm/pixel at the minimum distance of 2.1 cm) color close-up images of outcrops that provide textural and structural information on fine-scale geological features, such as mineralized veins and concretions, as well as valuable mineralogical information regarding the shape, cleavage, color, fluorescence and luster of individual mineral grains (Edgett *et al.*, 2012). The analysis of crystal morphology was conducted on MAHLI images using the Fiji distribution (Schindelin *et al.*, 2012) of the ImageJ open-source software (Abràmoff, Magalhães and Ram, 2004).

# 3. Geological setting of Vera Rubin ridge

## 3.1   Stratigraphy

The VRR structure was defined by its topographic prominence in orbital images. Following in-situ observations, it was later divided into two members representing the lower and upper sections of the ridge (Edgar et al., this issue). The lower member, named Pettegrove Point, is composed of well-cemented and finely laminated mudstones crosscut by light-toned filled-fractures. The upper member, Jura, conformably overlies Pettegrove Point and consists primarily of red blocky outcrops of finely laminated mudstone to fine-grained sandstones. The sedimentary facies and textures observed in both members of VRR are consistent with deposition in a lacustrine setting (Edgar et al., this issue). The lack of clear discontinuity between VRR and the underlying fluvial and lacustrine fine-grained sandstones and mudstones of the Murray formation suggests a continuity in the depositional environment. As such, VRR is regarded as a continuation of the Murray formation, which has been observed consistently along the rover traverse since sol ~750, when *Curiosity* reached the lowest strata of Mount Sharp at Pahrump Hills (encompassing >300 m in thickness) (Grotzinger *et al.*, 2015; Fedo *et al.*, 2018). We refer the reader to (Edgar et al., this issue) for a detailed description of the sedimentology and sedimentary rock textures.

Both Pettegrove Point and Jura members also display lateral color variations in the form of meter to decameter-scale patches of grey, fine-grained, laminated bedrock (Horgan et al., this issue), first noted on the flanks of Vera Rubin ridge during the approach (Figure 1), and which do not appear to be stratigraphically controlled (Edgar et al., this issue; Fraeman et al., this issue). The grey outcrops are more prominently observed in the Jura member where they display less resistance to mechanical erosion compared to adjacent red blocky outcrops. These outcrops form bright patches in shallow topographic depressions on top of the ridge (Fraeman et al., this issue) and were notably explored at the Jura, Rhynie, and Highfield localities (Figure 1). Since these lateral outcrop color and spectral variations appear to follow the geomorphology of VRR rather than its stratigraphy, they suggest a significant diagenetic overprinting on the formation of the ridge.

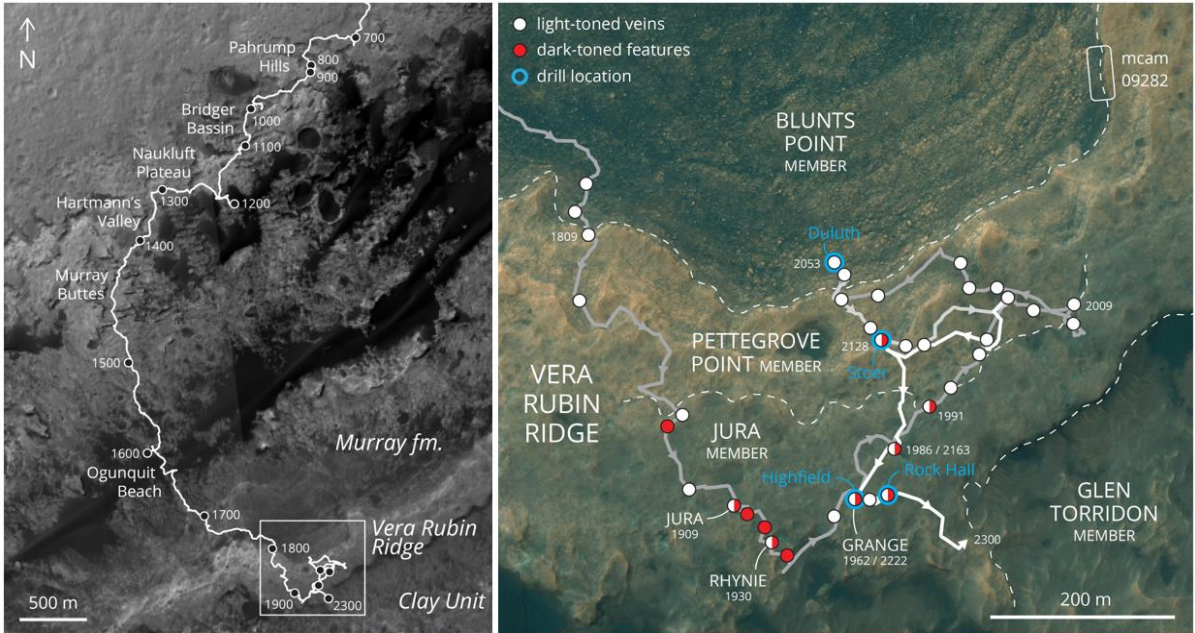
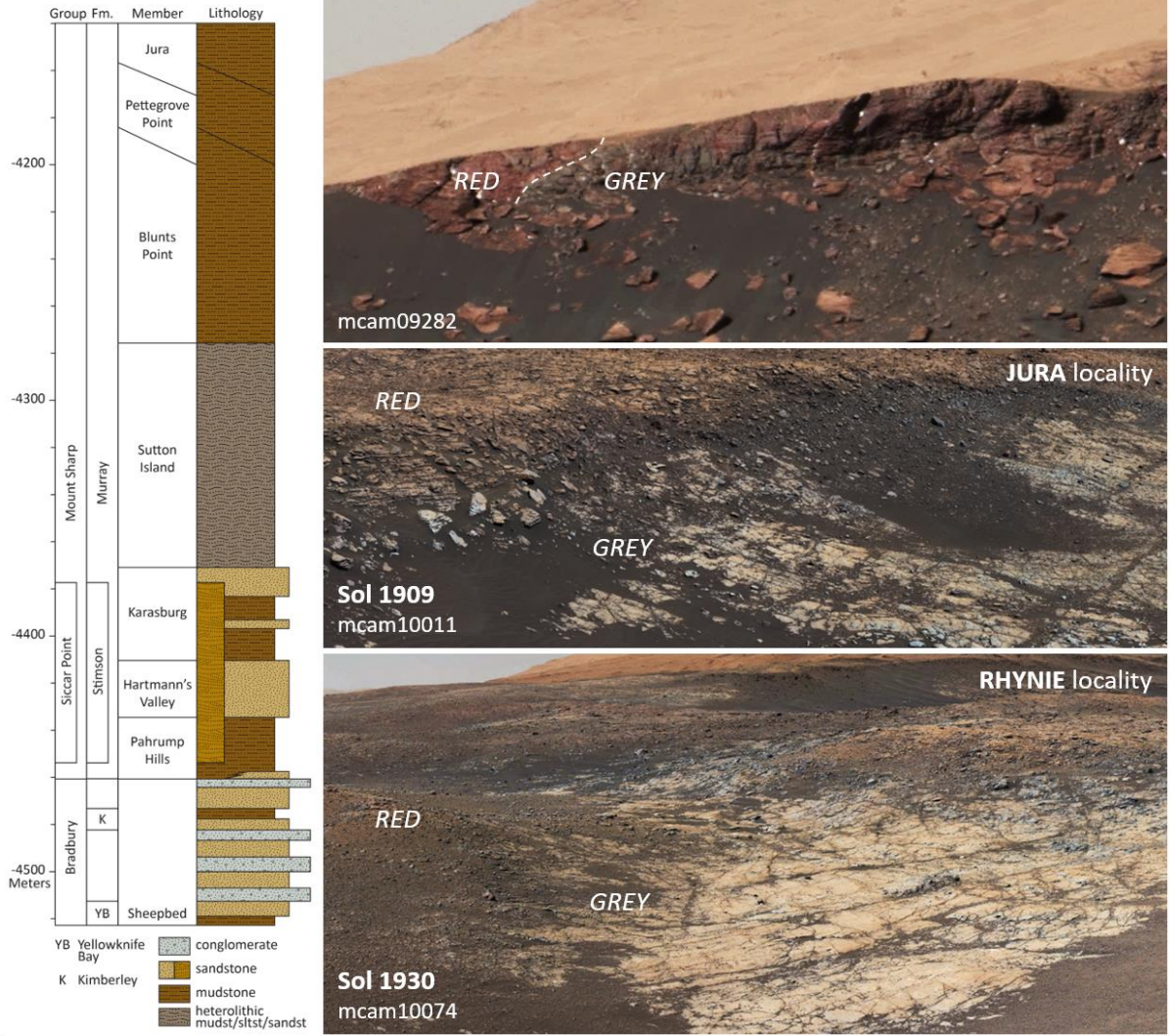

Figure 1: Location of Vera Rubin ridge along the rover traverse in Gale crater (top left), and a close-up view of the explored portion of the ridge displaying the location of ChemCam analyses of diagenetic features and of the drilling sites (top right). Stratigraphic column of the terrains explored by the rover (adapted from Edgar et al., this issue) showing that VRR is composed of the Pettegrove Point and Jura members and part of the Murray formation. Mastcam mosaics illustrating the lateral red and grey color variations in the bedrock on Vera Rubin ridge observed from a distance during the rover approach on the north flank of the ridge (mcam09282) and associated with topographic depressions, notably at the "Jura" (mcam10011), "Rhynie" (mcam10074) and "Highfield" localities.

### 3.2 Chemical and mineralogical composition

Chemical analysis from the ChemCam and Alpha-Particle X-ray Spectrometer (APXS) instruments onboard *Curiosity* report elemental bulk compositions for VRR bedrock within the compositional ranges of the Murray formation (Frydenvang et al., this issue; Thompson et al., this issue). The lack of enrichments in $FeO_T$ compared to the underlying Murray sedimentary rocks indicate that the iron forming the hematite detected from orbital data did not come from an external source. Nevertheless, ChemCam and APXS both report modest chemical variations in major elements with slightly higher silica and aluminum contents than in the underlying Murray rocks, as well as local depletion in iron. Minor elements also vary, with much lower lithium overall and significant variations in manganese which are not always correlated with variations in iron (David et al., this issue; Frydenvang et al., this issue; Thompson et al., this issue).

To further investigate this chemical variability, the *Curiosity* rover carried out three drilling campaigns on VRR—one in the Pettegrove Point member (Stoer drill target) and two in the Jura member, within both red outcrops (Rock Hall drill target) and grey outcrops (Highfield drill target) (Figure 1). Through the mineralogical analysis of the drill samples, the CheMin instrument (X-ray diffractometer) shows that hematite is detected at all three drill sites, albeit in lower abundance within the red outcrops of the Jura member (Rampe et al., this issue) compared to the rest of VRR. Overall, the VRR drill samples contain more hematite compared to the Murray formation (Rampe et al., this issue). In-situ Mastcam multispectral and ChemCam passive reflectance observations confirm that fine-grained red hematite is present systematically in the red to purple VRR outcrops, likely alongside with nanophase hematite (Horgan et al., this issue). Indeed, brushed VRR bedrocks display ferric spectral features (535 nm band depth, 750 to 840 nm slope, 860 nm band depth) (Fraeman *et al.*, 2018, 2019), with the exception of the grey outcrops of the Jura member that lack this ferric spectral signature (Horgan et al., this issue). This is interpreted to reflect a lack of fine-grained and nanophase hematite in the grey outcrops, but may still be consistent with the presence of coarser "grey" crystalline hematite (> 5-10 µm crystals) (Lane *et al.*, 2002). This hypothesis is supported by the detection of hematite by CheMin in both red and grey outcrops (Rampe et al., this issue). Lastly, CheMin observes a second, notable difference between the grey and red outcrops of the Jura member with the detection of jarosite and akaganeite within the only sampled red Jura outcrop (Rock Hall drill target), but not in a nearby grey outcrop (Highfield drill target) (Rampe et al., this issue). As such, these two diverging mineralogical observations between the grey and red outcrops will need to be carefully considered in any discussion regarding the diagenetic history of the Jura member.

In summary, stratigraphic, mineralogical, and chemical observations of the VRR outcrops suggest that various processes were involved at VRR, including diagenetic processes which are notably expressed as red to grey lateral color variations in outcrops. The results reported hereafter on small-scale diagenetic features, observed in both red and grey outcrops, provide further insights into the processes at play during the diagenesis.

## 3.3 Diagenesis in the context of Gale crater

The Curiosity rover has identified a range of mineral assemblages associated with diagenetic features, which are instrumental to reconstruct the post-depositional history of sedimentary deposits at Gale crater. Most notable amongst these features are the millimeter- to cm-wide light-toned Ca-sulfate veins observed throughout the traverse, across all stratigraphic members, which represent a late stage of diagenetic activity (e.g. Nachon *et al.*, 2014; L'Haridon *et al.*, 2018). In the Bradbury formation, at Yellowknife Bay (Figure 1), the light-toned veins are composed of nearly pure Ca-sulfate, and have therefore been interpreted as precipitation from a fluid that dissolved pre-existing, more impure Ca-sulfate deposits (Schwenzer et al., 2016). Within the mudstones of the Murray formation (Figure 1), the light-toned veins are more pervasive and were interpreted to have formed by hydraulic fracturing (Caswell and Milliken, 2017; De Toffoli *et al.*, 2020). There, the light-toned Ca-sulfate veins are locally associated with Mg-sulfates, Fe-sulfates, fluorite (Nachon *et al.*, 2017), as well as P- and Mn-rich mineral phases (Meslin *et al.*, 2018). In addition, large silica-rich haloes are observed along fractures crossing through the Murray formation and the overlying eolian sandstone of the Stimson formation (Figure 1) (Frydenvang *et al.*, 2017).

The variability in compositions and textures of the diagenetic features and their distribution across the Murray formation attest to a complex history during diagenesis, with local occurrences (e.g., fluorite detections) associated with more pervasive features (Ca-sulfate veins). There, diagenesis involved multiple stages of groundwater circulation with varying fluid compositions that are still only partly understood. In this context, the formation of hematite in association with light-toned Ca-sulfate veins locally at VRR constitutes an important piece of the puzzle, and raises questions with respect to the specificity of VRR within the Murray formation.

# 4. Results

In order to contextualize ChemCam chemical and spectral observations, each type of small-scale diagenetic feature encountered on the ridge will first be described thoroughly with the help of both RMI and MAHLI images (section 4.1). However, as the textures and morphologies of the diagenetic features directly relate to their chemical (and mineralogical) compositions, the descriptions will also introduce key observations on chemical composition, which will be further detailed in the subsequent sections (sections 4.2 and 4.3).

## 4.1 Description and major element chemistry of diagenetic features

### 4.1.1 Light-toned veins

Pervasive to all geological terrains explored so far by the *Curiosity* rover, millimeter- to centimeter-scale light-toned mineralized veins are consistently observed across VRR. The Pettegrove Point member is dominated by curvi-planar light-toned veins and erosion resistant fins which transition upwards to more recessive fine veins and fracture fills in the Jura member (Bennett *et al.*, this issue); all clearly diagenetic since they disrupt primary sedimentary structures (i.e. sub-horizontal bedding, lamination, cross-bedding). In addition, the morphology of the fracture network exhibits no clear preference for orthogonal intersections, with light-toned fractures that often "wander" and isolate host rock fragments. Similar morphologies were also observed in the terrains below the ridge, where light-toned veins are interpreted to have formed by hydraulic fracturing, induced by pore-fluid pressure build-up in a compacting sedimentary basin (Caswell and Milliken, 2017; Kronyak *et al.*, 2019; De Toffoli et al., 2020). The composition of the light-toned veins is strongly enriched in sulfur and calcium, thus consistent with Ca-sulfate mineralogy (c.f. section 3.2), in line with the previous observations on similar features along the rover traverse (Nachon et al., 2014; Rapin et al., 2016).

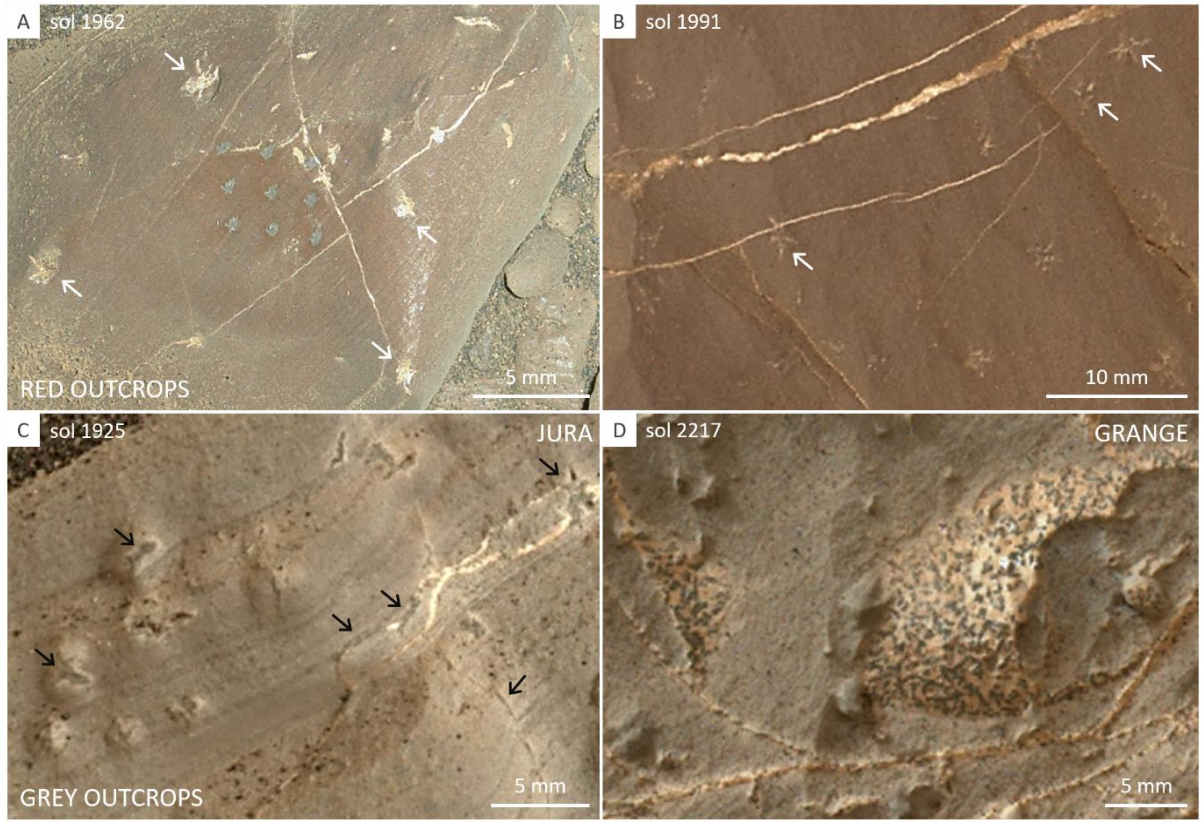

Figure 2: MAHLI observations of light-toned veins in the Jura member of VRR. In red outcrops: light-toned veins crosscutting light-toned crystals (white arrows) with shapes consistent with gypsum habit (A, B), which in some locations also incorporate hints of darker-toned material (B). In grey outcrops: dark-toned material observed as partial fills in light-toned veins and in angular crystal casts (black arrows) (C), as well as polygonal inclusions within "wandering" light-toned veins (D). MAHLI images: 1962MH0001630000704514R00 (A), 1991MH0001900010800121C00 (B), 1926MH0003690000703349R00 (C), and 2217MH0007060010802994C00 (D).

Lateral variations of vein texture and composition are notably observed in the Jura member. In the red outcrops, the thin straight fracture-filling veins are recessive with a smooth texture and uniformly white color (Figure 2A-B). In contrast, in the grey outcrops, the veins *Curiosity* encountered display a rugged texture, often encompassing dark-toned polygonal inclusions and partial vein fills (c.f. section 4.1.3 and Figure 2C-D).

Lastly, *Curiosity* observed large light-toned veins within the grey outcrops that encompass grey-toned diffuse patches where ChemCam measured high $FeO_T$ content in addition to their Ca-sulfate component (Figure 3). These observations are reminiscent of Fe-enrichments observed within light-toned Ca-sulfate veins lower in the Murray formation (L'Haridon et al., 2018).

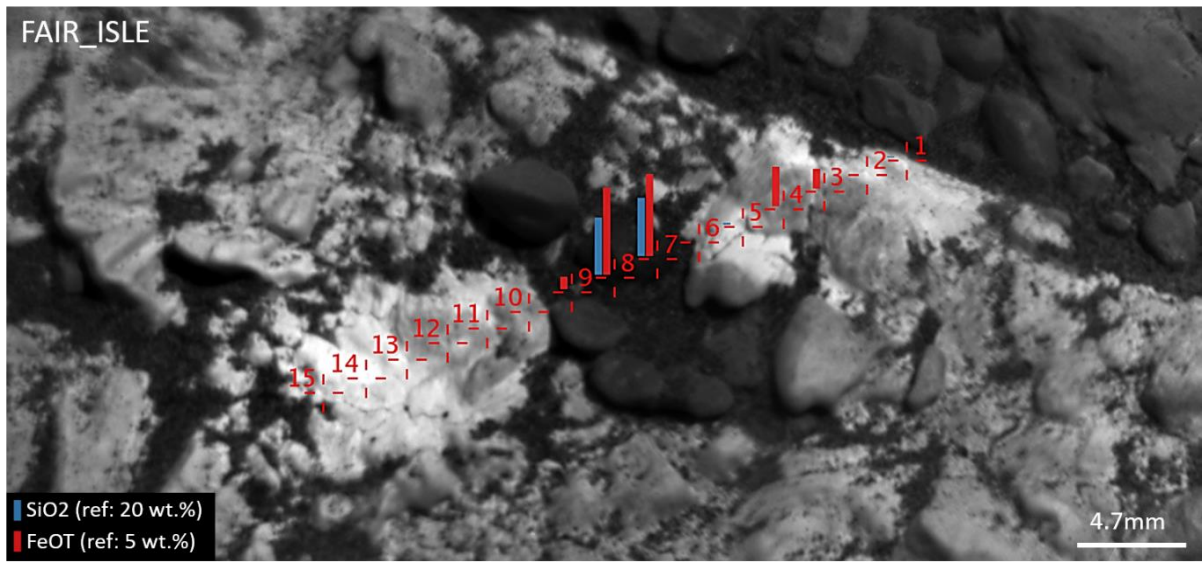

Figure 3: ChemCam analyses on a large light-toned Ca-sulfate vein (Fair_Isle target) show diffuse Fe-enrichment associated with greyer material within the vein (around points #3 and #4), in the grey outcrops of the Jura locality. The large variations in composition on locations #7 and #8 reflect the sampling of wind-blown sand, so these two points will not be considered in the discussion.

### 4.1.2 Angular crystal casts

Both red and grey outcrops host angular, millimeter-scale crystal casts (e.g., Figure 2A-B-C and Figure 4) (Bennett *et al.*, this issue). While individual casts are observed as euhedral tabular crystal shapes, they commonly coalesce into V-shaped (or "swallowtail") and star-shaped crystalline aggregates (Figure 2A-B-C and Figure 4). Though crystal twinnings were not identified with absolute certainty in the VRR outcrops, the observed geometry could be consistent with the crystal habit of gypsum (El-Tabakh, Riccioni and Schreiber, 1997; Paik *et al.*, 2007; Warren, 2016).

In the red outcrops, crystal casts exhibit a light-toned color, strikingly similar to the adjacent light-toned veins (Figure 2A-B), suggesting a similar Ca-sulfate composition which would be consistent with the inferred gypsum crystal habit. Unfortunately, the light-toned crystal casts observed in the red outcrops were not successfully sampled by the ChemCam instrument due to ChemCam targeting uncertainties and their small size. The light-toned crystals observed in some red outcrops also exhibited hints of replacement by a darker-toned material (Figure 2B). In contrast, the crystal-shaped casts encountered in the grey outcrops are mostly expressed as void spaces (crystal molds) or comprise mainly dark-toned material with hints of residual light-toned material (Figure 2C and Figure 4, the residual light-toned material is notably visible in inset B). ChemCam was more successful in targeting these dark-toned angular crystal casts, reporting a composition enriched in $FeO_T$ (Figure 4).

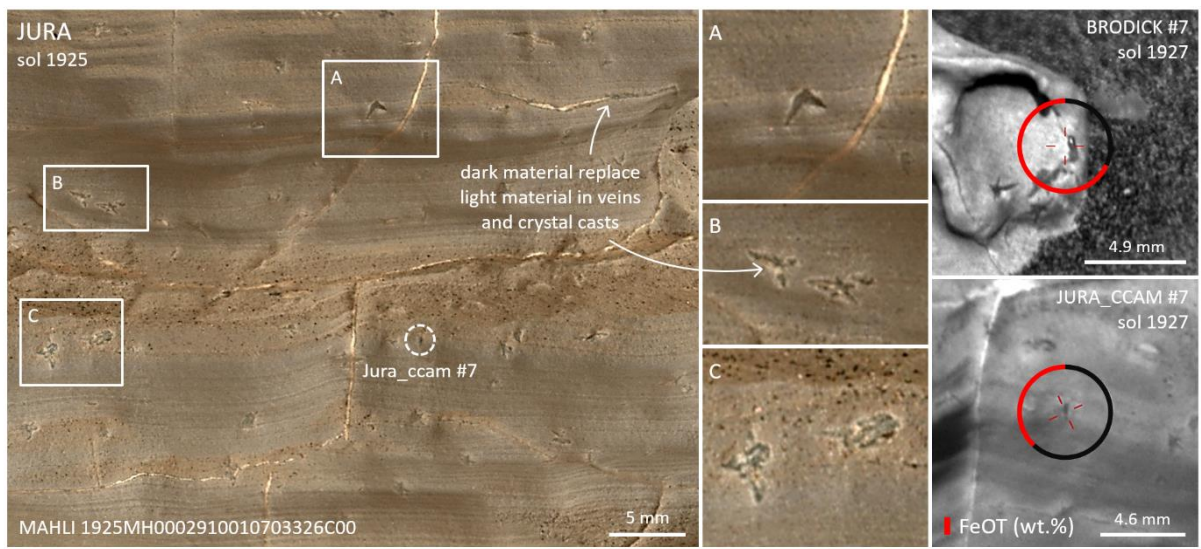

Figure 4: Crystal shapes and molds observed on sol 1927 ("Jura" locality, within grey outcrops) showing swallowtail-like shapes (inset A) as well as star-shaped aggregates (insets B and C) reminiscent of gypsum crystal habit. These features were sampled by ChemCam on the Brodick #7 and Jura_ccam #7 targets and show enrichment in $FeO_T$ (59.6 and 34.5 wt.% respectively) compared to the host rock (~20 wt.%).

The euhedral crystal casts appear to be randomly distributed in the host rocks, with no preferred orientation with respect to bedding planes (e.g. Figure 4). Additionally, at the MAHLI image resolution, the casts do not appear to disrupt primary sedimentary structures (e.g. laminations on Figure 2A-C and Figure 4), which would have indicated displacive mineral growth within unconsolidated sediments (Murray, 1964; Hardie, 1985; Warren, 2016).

### 4.1.3 Dark-toned polygonal features and fracture fills

The grey outcrops also host dark-toned nodular features (Figure 5) that are characterized by high $FeO_T$ abundances (see 3.2). These features are preferentially encountered along light-toned veins, as inclusions and fracture-fills (Figure 2C-D and Figure 5), but are also observed scattered in the adjacent host rock (e.g., Figure 4 and Figure 5A-B-C) displaying angular shapes similar to the light-toned crystals observed in the red outcrops (c.f. section 4.1.2). Interestingly, the host rock in close proximity to some of the largest dark-toned features is lighter-toned (or bleached) over 1 cm (e.g. around the larger nodular dark-toned features and adjacent fracture-fills; Figure 5C-D) with lower $FeO_T$ content compared to the surrounding bedrock.

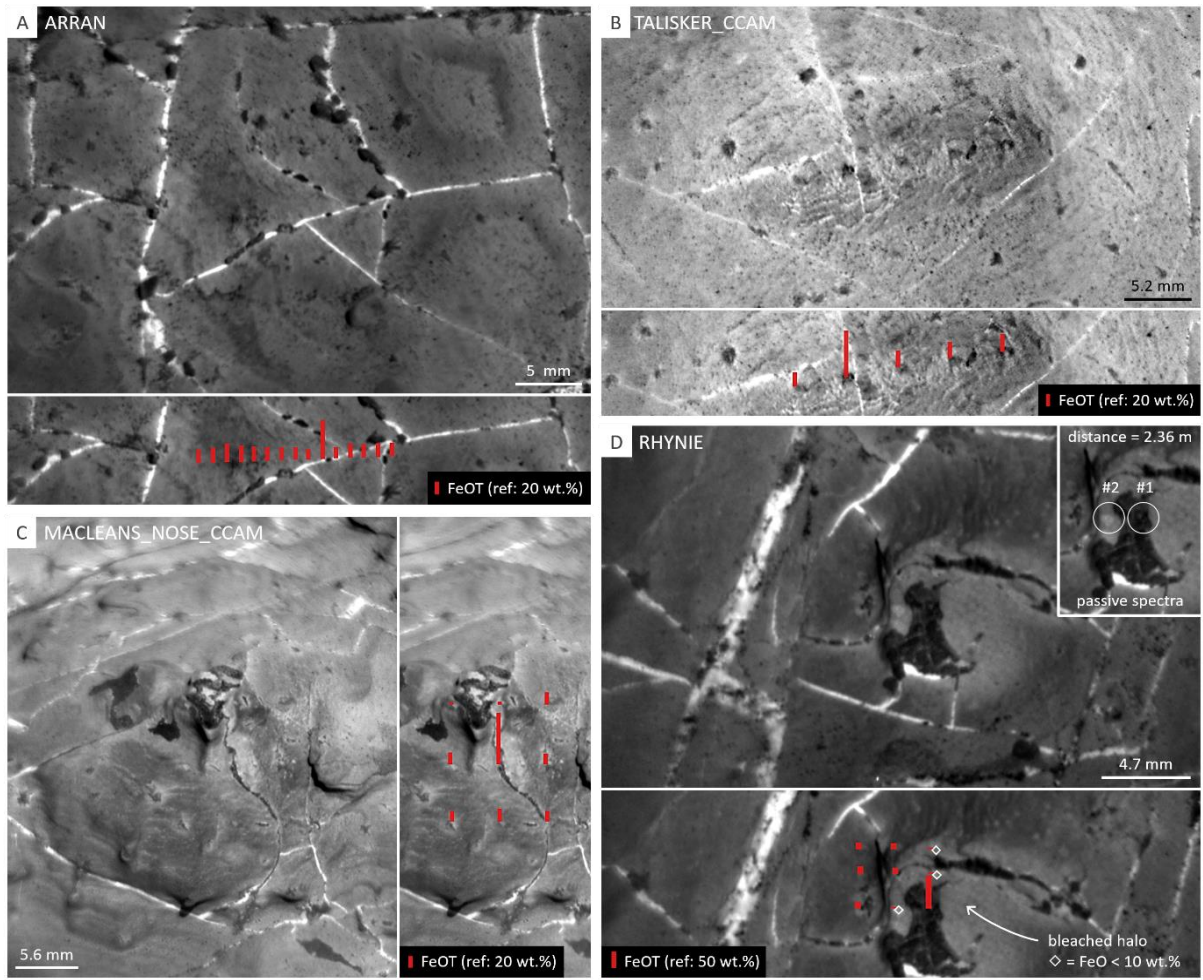

Figure 5: ChemCam observations of dark-toned features, encountered preferentially along light-toned Ca-sulfate veins (e.g. A-B). Nodular dark-toned features are also observed (e.g. C-D) forming connected diagenetic complexes with dark-toned and light-toned veins. Dark-toned features are encountered as inclusions or partial fills in light-toned veins but are not intersected or crosscut by these veins. Dark-toned angular casts are also observed (white arrows). The host rock near the larger dark-toned features appears lighter-toned (bleached halos, e.g. D). $FeO_T$ relative content for each ChemCam sample location is indicated to illustrate the association of high-Fe with dark-toned features and low-Fe with bleached light-toned halos (the red bar in the legend indicates the reference, set at 20 wt.% for A-C and 50 wt.% for D). The field of view for passive observations is indicated on the Rhynie target, for point #1 (dark-toned Fe-rich feature) and point #2 (low-Fe bleached light-toned halo). RMI images for B & D were colorized using context Mastcam images.

Fracture fills transition laterally from light-toned to dark-toned material within the same fracture (e.g. Figure 2C and Figure 5C-D) and light-toned veins never crosscut the dark-toned features (Figure 2C and Figure 5). Furthermore, the dark-toned inclusions observed within the "Grange" target, a light-toned vein (Figure 2D and Figure 6), display euhedral crystal shapes. The dark-toned inclusions display a hexagonal crystal habit, as denoted by the elongated hexagonal prism and regular hexagon shapes associated with the dark-toned euhedral crystals within the light-toned "Grange" vein (Figure 6, inset A and B). In addition, the angles measured on unambiguous dark-toned crystals within the vein (n = 600) show a distribution with a median at ~120° (Figure 6), albeit with an elevated number of angles near 90°. Such a distribution of angles suggests that the dark-toned mineral pertains to the hexagonal crystal family (three axes of symmetry with an included angle of 120°, and a fourth axis perpendicular to the others).

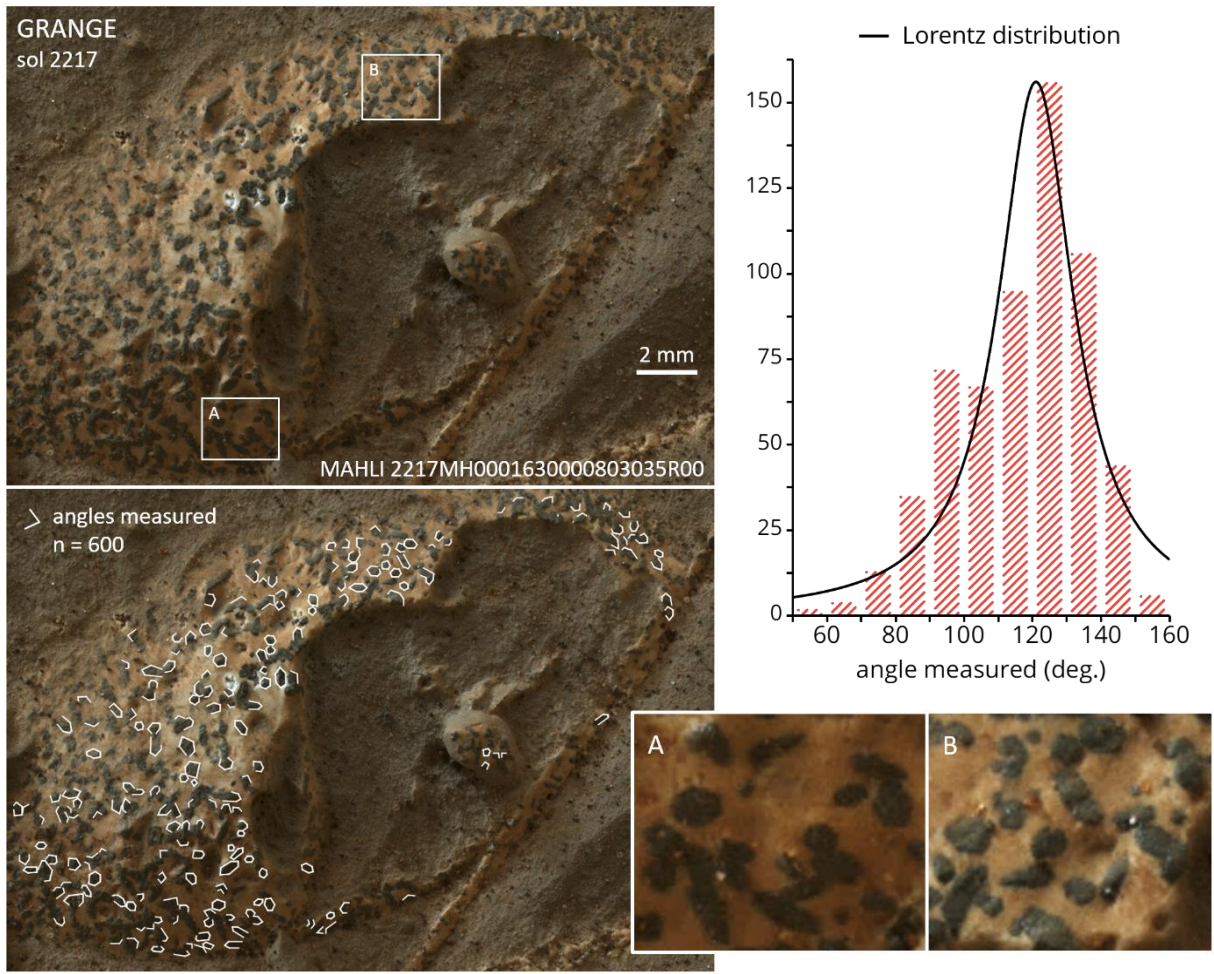

Figure 6: Morphological analysis of dark-toned crystals in the Grange target (left). The angles measured on identifiable crystals (white lines, bottom left) show a distribution centered around 120° (right). This distribution of angles would be consistent with a hexagonal crystal family and reflects the numerous hexagonal shapes observed on the dark-toned crystals (i.e. insets A and B).

Lastly, *Curiosity* observed dark-toned polygonal and elongated features that protrude from the grey outcrops in positive relief, along fracture planes, reflecting a higher resistance to erosion than the surrounding host rock. ChemCam sampled only once these erosion-resistant features—which displayed elevated $FeO_T$ (Haroldswick #17 inset, Figure 7)—although the feature appears to have been blasted by the successive laser shots. The morphological similarities with neighboring dark-toned features (notably polygonal fracture-fills) and with adjacent polygonal molds observed along fractures (e.g. Islay target, Figure 7) suggest a genetic link between these features (Bennett *et al.*, this issue).

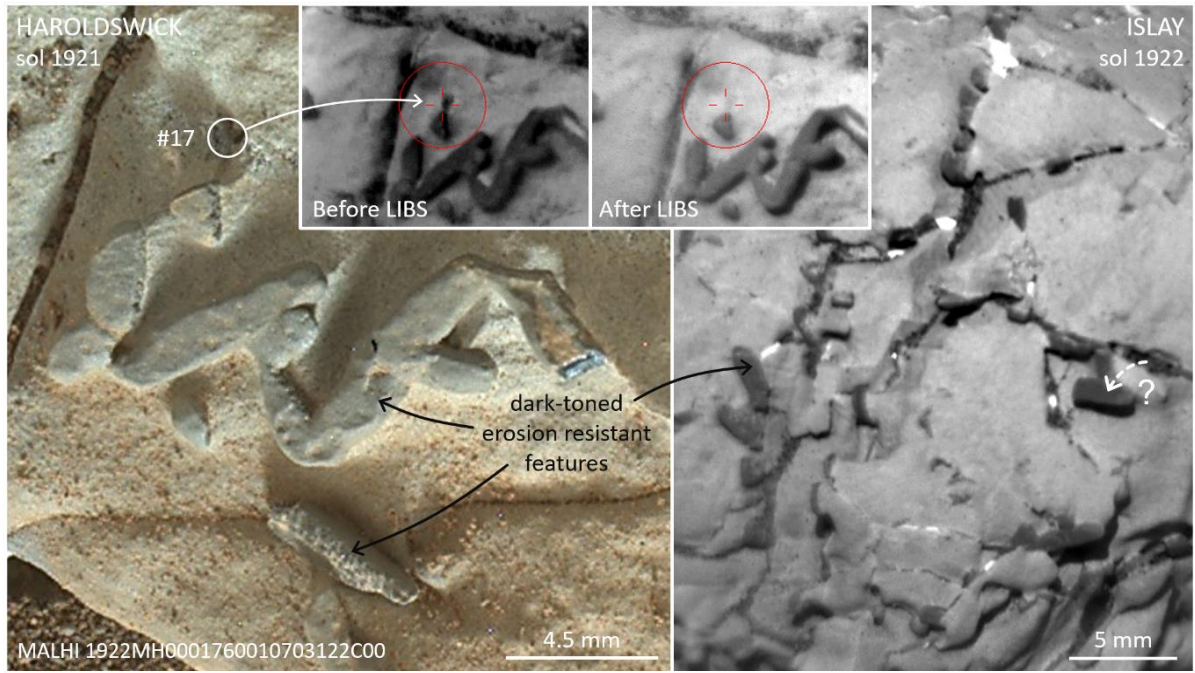

Figure 7: Polygonal and elongated erosion-resistant dark-toned features ("stick-shaped") observed at the "Jura" locality within grey outcrops, which show morphological similarities with high-Fe dark-toned features (notably fracture-fills) and potentially associated with polygonal molds observed along fractures (e.g. Islay target). ChemCam may have sampled a dark-toned erosion-resistant feature (inset; Haroldswick #17), which showed elevated Fe abundances (~39 wt.%) but was blasted by the ChemCam laser shots.

## 4.2 Detailed chemistry (major and minor elements) of diagenetic features

The chemical abundances from ChemCam are reported in ternary diagrams in Figure 8 (in molar proportion, based on quantified oxide weight percent reported by ChemCam), illustrating the chemical trends between Fe and Si with respect to Al, Mg and Ca (Figure 8A-C-D) and between Fe, Si+Al and Mg+Na+K (Figure 8B). This dataset includes observations on host rock (3175 points), light-toned veins (207), dark-toned features (101), and bleached host rock (71) (see section 2.2).

The dark-toned features, including dark-toned material within angular crystal casts, show a composition dominated by enrichment in Fe and corresponding decreases in other major elements (notably Si, Al, Mg, K, and Na) compared to the host rock compositions (Figure 5 and Figure 8). High-Fe observations are not associated with the detection of volatile elements (S, Cl, P, F or C; Figure S-1), ruling out any Fe-sulfates or Fe-phosphates (Forni *et al.*, 2015; Nachon *et al.*, 2016; Meslin *et al.*, 2018). The H emission line is also relatively low (Figure 9), thus pointing toward poorly hydrated Fe-oxide minerals.

The only exception was observed on St Cyrus #8, (Figure S-1) where high Fe is associated with Cl and high H, and which is interpreted as a detection of akaganeite [FeO(OH,Cl)]. That observation was however not clearly associated with an identifiable dark-toned feature (Figure S-1), and thus interpreted to reflect the presence of akaganeite in the host rock, which was also observed by the CheMin instrument at the Rock Hall drill site (Rampe et al., this issue).

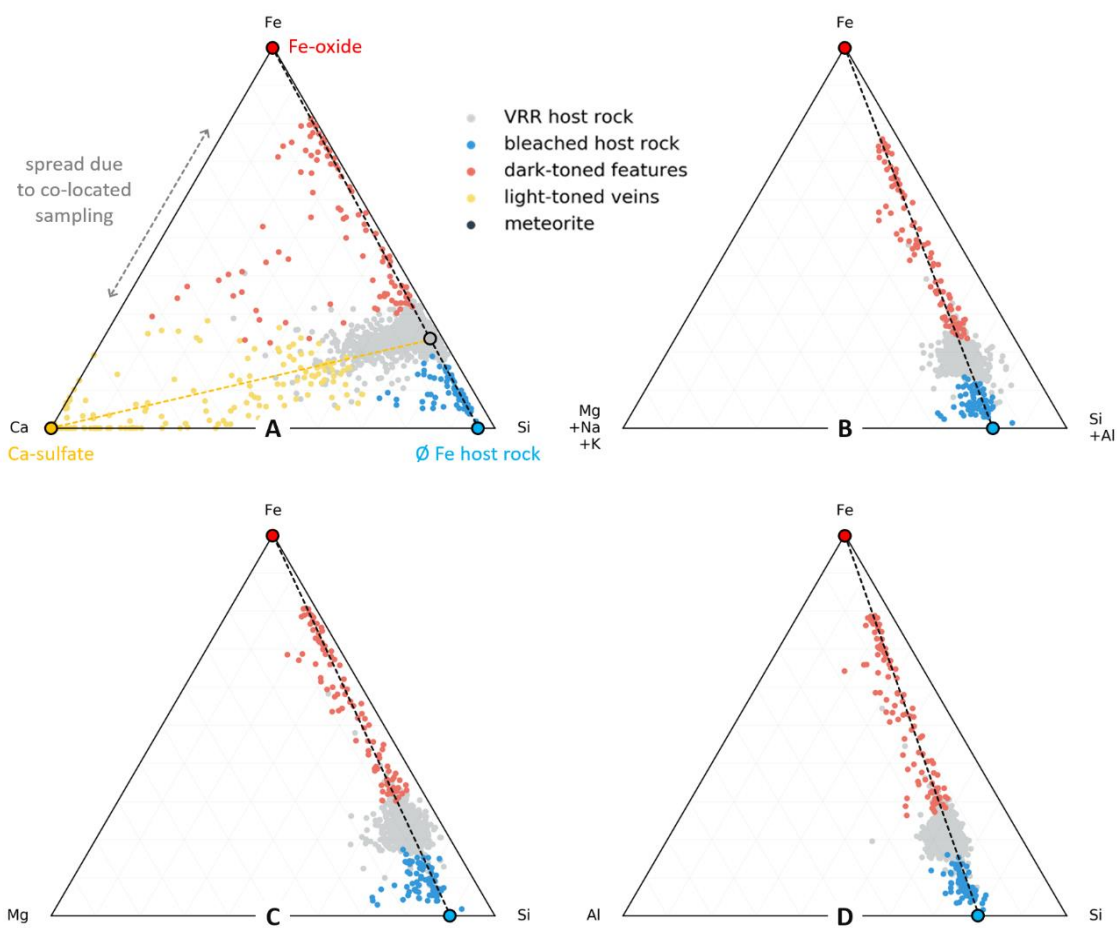

Figure 8: Ternary diagrams showing ChemCam quantified abundances (molar percentages) for major elements: dark-toned features (in red) are associated with very high Fe content, trending toward a pure Fe composition, coupled with low-Fe observations on bleached light-toned halos. Diagram (A) shows a spread toward Ca for dark-toned features due to the contribution of Ca-sulfate from adjacent light-toned veins, and sampling at the interface between both features. Diagrams (B), (C) and (D) show that the low-Fe content in bleached host rock targets is not associated with a loss in any other major elements and as such show mobilization of Fe from the host rock.

The trend in major elements shows an enhanced Fe abundance on dark-toned features, scattered along a mixing line between the host rock compositions and the Fe apex of the diagram, due to the small size of these features (laser shots on diagenetic features also include some of the surrounding host rock). The composition of these diagenetic features trends toward a pure Fe-oxide end-member composition (Figure 8).

The spread observed toward elevated Ca in some of these features (Figure 8A) is attributed to the sampling of both dark-toned features and light-toned veins by the same ChemCam observation point, due to their close spatial association (Figures 4-5). Indeed, light-toned veins show a composition consistent with a Ca-sulfate dominated composition (Figure 8A), with high Ca content associated with a lower sum of oxides and S detection (Nachon *et al.*, 2014); interpreted as bassanite (CaSO$_4$ x 0.5 H$_2$O) based on the H content detected by ChemCam (Rapin *et al.*, 2016). In addition, some large veins also encompass grey patches that display high Fe in association with Ca-sulfate (c.f. section 4.1.1 and Figure 3).

The lighter-toned bleached halos observed in the host rock surrounding several high-Fe dark-toned features (e.g. Rhynie #2, Figure 5) show depleted Fe content compared to other VRR host rock observations (commonly < 10 wt.% compared to 18-20 wt.% typically observed in the host rocks, Figure 5C-D and 8). Interestingly, lower FeO$_T$ abundances are not correlated with depletion in any other major elements, which preserve the same relative abundances (Figure 8). As such, these observations suggest preferential leaching of Fe in these bleached halos compared to the adjacent host rock composition.

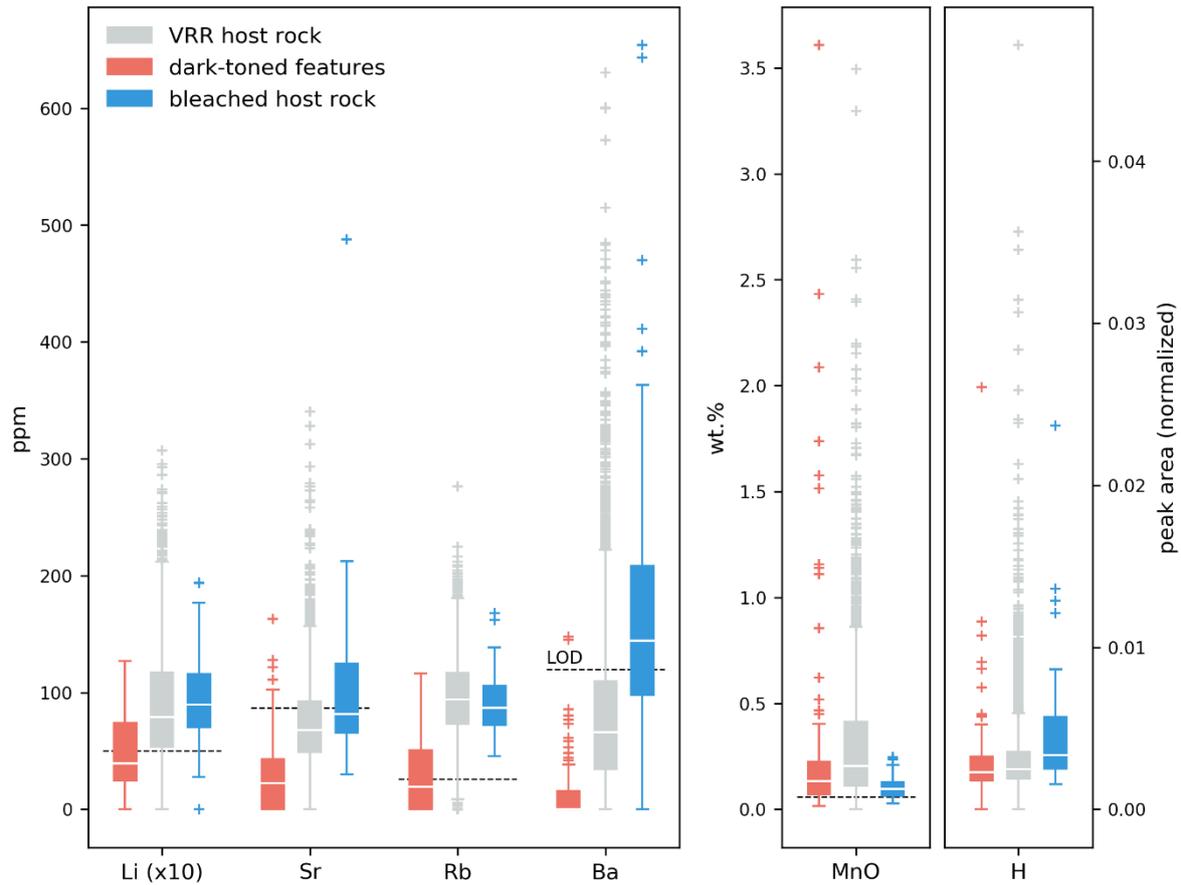

Figure 9: Box plots representing quantified abundances for Li, Rb, Sr, Ba, and MnO, as well as H normalized peak area (c.f. Rapin *et al.*, 2017), in dark-toned Fe-rich features (red), VRR host rocks (grey), and low-Fe bleached host rocks (blue). Abundances in Ba, Rb, Sr, and Li for bleached low-Fe targets are in range with typical host rock values (although slightly enriched in Ba and Sr) but are depleted in dark-toned Fe-rich features. However, MnO abundance in dark-toned Fe-rich features is in line with host rock values and appears depleted in low-Fe bleached targets. H appears to be in line with the low host rock values in dark-toned Fe-rich features but slightly enriched in low-Fe bleached targets. Limits of detection (LOD) are indicated for Li, Sr, Rb, and Ba (V. Payré *et al.*, 2017), as well as for MnO (Lanza *et al.*, 2014).

Minor element abundances reported by ChemCam show mostly depleted contents in dark-toned Fe-rich features for Ba, Rb, Sr, and Li compared to VRR host rock values, whereas low-Fe bleached halo abundances for the same elements are within range of host rock values, except for a slight enrichment in Ba and Sr (Figure 9). In contrast, MnO abundances in dark-toned features are within the range of host rock values but depleted in low-Fe bleached halos (Figure 9), showing that the two redox-sensitive elements Fe and Mn are depleted in concert, while other elements are in stable abundance.

## 4.3 Passive spectra observations

ChemCam passive spectral analyses on the various rocks of the ridge reveal variability in the strength of the spectral features diagnostic of ferric iron (i.e. absorptions near 535 and 670 nm and a downturn after ~750 nm). Red and purple VRR host rocks show a ferric signature in passive reflectance spectra, including the red outcrops of the Jura member (Figure 10), consistent with the presence of ferric phases, and in particular hematite (Fraeman *et al.*, 2013; Johnson *et al.*, 2017; Jacob et al., this issue).

In contrast, the passive spectra observed in the grey outcrops are much flatter in the 600-800 nm spectral region (Figure 10). This observation is also consistent with Mastcam multispectral observations of grey outcrops and a few resolved dark diagenetic features, which also show relatively flat spectra without clear absorption bands, even at longer wavelengths not detectable by ChemCam (Horgan et al., this issue). Coarse-grained and crystalline "grey" hematite would be consistent with such passive spectral observations (c.f. USGS and RELAB reference spectra, Figure 10) since grain size, contaminants and crystallinity can also impact spectral ferric features (Morris *et al.*, 1985; Lane *et al.*, 2002; Johnson *et al.*, 2015). The spectral variations observed by ChemCam and Mastcam may thus reflect the differences in mineralogy observed by CheMin (Rampe et al., this issue; Horgan et al., this issue), notably concerning the Fe-oxide components. The "red" hematite [$\alpha$-Fe$^{3+}_2$O$_3$] characterized by its ferric spectral signature in the red outcrops is giving way to "grey" crystalline hematite [$\alpha$-Fe$^{3+}_2$O$_3$] associated with a subdued spectral signature in the grey outcrops (Rampe et al., this issue). Low-Fe bleached halos do not show any spectral evidence for the presence of ferric iron, in line with the observations on the adjacent grey host rocks (Figure 10). Similarly, the relatively flat passive reflectance spectra (i.e. lacking ferric absorption features) associated with dark-toned Fe-oxide features suggest the presence of either mixed-valence Fe-oxide (i.e. magnetite) or coarse-grained grey hematite (Figure 10).

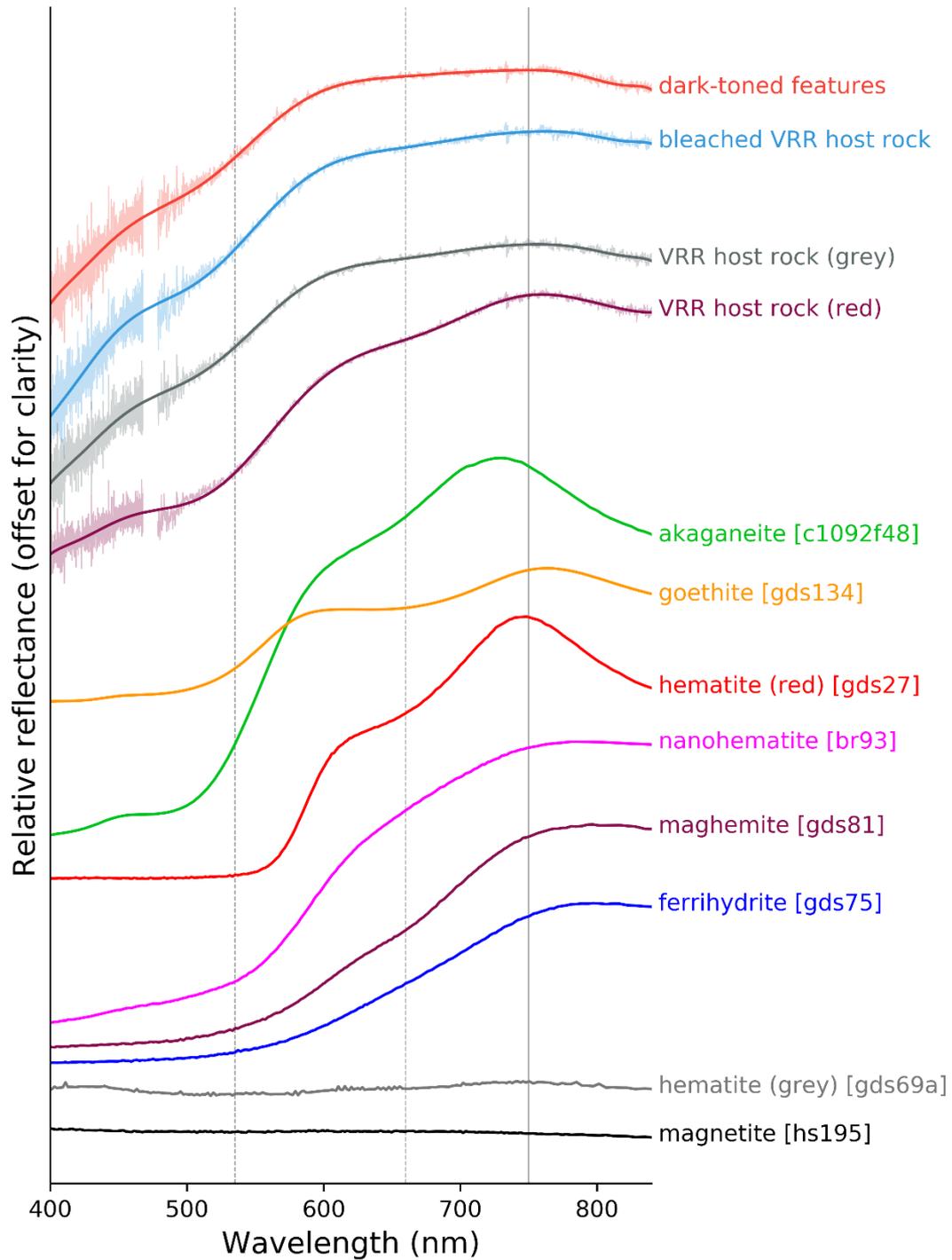

Figure 10: Passive reflectance spectra of high-Fe dark-toned features (in red, Rhynie #1), low-Fe bleached halo (in blue, Rhynie #2), as well as grey (in grey, Ben Loyal #5) and red (in purple, Holyrood #1) host rock targets. The solid line overprinted on spectral data corresponds to a 15th order polynomial fit, shown here for clarity. High-Fe dark-toned features, low-Fe bleached halos, and grey host rocks lack adsorption bands near 535 nm and 670 nm (dashed grey lines) and a downturn after 750 nm (solid grey line) consistent with $Fe^{3+}$ phases observed on red host rocks. Library spectra for hematite [$\alpha$-$Fe_2O_3$] (red, grey and nanophase), maghemite [$\gamma$-$Fe_2O_3$], magnetite [$Fe^{2+}Fe^{3+}_2O_4$], ferrihydrite [$Fe^{3+}_2O_3 \times 0.5H_2O$], goethite [$\alpha$-FeO(OH)] and lepidocrocite [$\gamma$-FeO(OH)] from the USGS database, as well as akaganeite [$Fe^{3+}O(OH,Cl)$] from the RELAB database, are provided for comparison. The laboratory spectra are scaled at half intensity, and all spectra are offset for clarity.

## 4.4 Summary of Observations

Examination of the texture and composition of small diagenetic features at VRR enables an exhaustive classification (Table 1). Overall, the light-toned veins are composed of Ca-sulfate with local and discrete enrichments in $FeO_T$. The abundant dark-toned features are present throughout the grey outcrops as neo-formed crystals or mineral pseudomorphs and along the light-toned veins as partial fracture fills and nodular inclusions. These features display a unique composition consistent with anhydrous iron oxides. The passive reflectance observations lack a ferric spectral signature, which enables us to discard the presence of "red" hematite but does not help to discriminate between crystalline hematite and magnetite. Nevertheless, the hexagonal habit observed on Fe-oxide dark-toned crystals within light-toned veins (Figure 6) provides insights into the crystalline structure of the observed Fe-oxide, which would be consistent with grey hematite, but not with magnetite. Indeed, hematite crystallizes in the trigonal system (part of the hexagonal family) and is commonly observed as hexagonal platelet-type crystals (Sugimoto *et al.*, 1993, 1996; Cornell and Schwertmann, 2003), while magnetite crystallizes in the cubic system and almost exclusively form octahedral crystals (Cornell and Schwertmann, 2003).

Diagenetic mineral replacements (pseudomorphism) appear to be limited to the grey outcrops, as the light-toned crystal casts observed in the red outcrops only show very limited evidence for dissolution and/or mineral replacement (c.f. Figure 2B). So are dark-toned fracture fills, which are scarcely observed outside of the grey outcrops. The bleached halos observed in the host rock surrounding some of the larger hematite-rich dark-toned features may give insights into the diagenetic processes that occurred in the grey outcrops. The high-$FeO_T$ and low-$FeO_T$ observations (respectively dark-toned features and bleached halos) neatly plot along a mixing-line between Fe-oxide and "Fe-free" host rock compositions, suggesting that Fe was mobilized locally from the host rock to form the Fe-oxide dark-toned features during diagenesis. Fe and Mn appear to exhibit a similar behavior, which would be consistent with the involvement of redox-driven chemical processes in the formation of these features since Mn mobility also greatly depends on its oxidation state (e.g., Post, 1999). Higher temperatures may also enhance the mobility of these elements, as it was suggested from CheMin analyses (Rampe et al., this issue), but the hydration of Ca-sulfates observed in light-toned veins (mostly basanite) indicates that temperatures did not exceed 60°C at the time of their formation (Rapin *et al.*, 2016, and references therein). This observation, together with the lack of mobility in other elements, discard a high-temperature scenario for the later stages of diagenesis at least, which are associated with the formation of hematite dark-toned features and crystals alongside Ca-sulfate veins.

The morphological similarities between all these neighboring dark-toned features suggest a genetic link between them. Observations of mineral pseudomorphs suggest that hematite replaced the original light-toned mineral (likely gypsum) during diagenesis—preserving its initial crystal shape through dissolution (formation of empty crystal molds) and re-crystallization (e.g. Lougheed, 1983; Paik *et al.*, 2007). The observations of pseudomorphism and low-Fe bleached halos around the dark-toned features suggest chemical interactions and Fe mobility within the host rock at the time of their formation. In contrast, the presence of dark-toned partial fracture-fills and dark-toned inclusions in the Ca-sulfate veins indicates a relationship with the stage of fracturing and vein formation during diagenesis. Accordingly, the observations of euhedral dark-toned Fe-oxides within a large Ca-sulfate vein at "Grange" indicate that the dark-toned inclusions and the light-toned veins formed contemporaneously, which constitutes a critical observation in order to determine the chronology of the diagenetic episodes (section 5.1) as well as mineral parageneses (section 5.2).

Table 1: Summary of diagenetic observations on Vera Rubin ridge

| | Feature | Description (section 4.1) | Composition (section 4.2, 4.3) |
|---|---|---|---|
| **RED OUTCROPS** | Light-toned veins | – linear and thin morphology<br>– smooth texture, uniform white color<br>– crosscut primary sedimentary structures and angular crystal casts | – high Ca and S, variable Sr<br>– H consistent with bassanite |
| | Angular crystal casts | – euhedral and tabular crystal shapes<br>– form V-shaped and star-shaped aggregates<br>– uniform white color (similar to light-toned veins)<br>– primary sedimentary structures not deformed<br>– exhibits hints of replacement by darker-toned material in some outcrops | – no data<br>(inferred to be Ca-sulfate) |
| **GREY OUCROPS** | Light-toned veins | – rugged texture<br>– encompass dark-toned inclusions, partial fills, and diffuse grey patches<br>– crosscut primary structures, but not dark-toned features or dark-toned pseudomorphs | – high Ca and S, variable Sr<br>H consistent with bassanite<br>– high Fe for dark-toned inclusions, partial fills and grey patches |
| | Angular crystal casts | – euhedral and tabular crystal shapes<br>– form V-shaped and star-shaped aggregates<br>– occasionally observed as crystal molds<br>– substitution of precursor light-toned material by dark-toned material (pseudomorphism)<br>– primary sedimentary structures not deformed | – elevated Fe compared to host rock (dark-toned material) |
| | Dark-toned features, fracture-fills, and erosion-resistant casts | – nodular features observed in association with light-toned veins (inclusions and fracture-fills)<br>– polygonal and elongated casts resistant to mechanical erosion, possibly match adjacent polygonal molds along fractures<br>– dark-toned inclusions display hexagonal crystal shapes, possibly pertaining to the hexagonal family (distribution of angles around 120°) | – very high Fe, low H<br>– Mn at a similar level compared to host rock composition<br>– depleted in other elements<br>– lack ferric absorption bands |
| | Bleached halos | – lighter-toned halos in the host rock<br>– observed around the larger dark-toned nodular features and facture-fills | – low Fe and low Mn<br>– other elements in line with the host rock composition, except for elevated Ba<br>– lack ferric absorption bands |

# 5. Discussion

## 5.1 Diagenetic history of Vera Rubin ridge

The VRR sedimentary rocks, deposited in a lacustrine environment, record a complex post-depositional geological history, as evidenced by the variety of small-scale diagenetic features observed on the ridge (Table 1). Crosscutting relationships between the observed diagenetic features, and how they affect the primary sedimentary structures, provide insights into their relative timings of formation. Thus, we discuss here the formation of the geological features observed on VRR, to reconstruct the sequence of events that occurred during its diagenetic history (Figure 11).

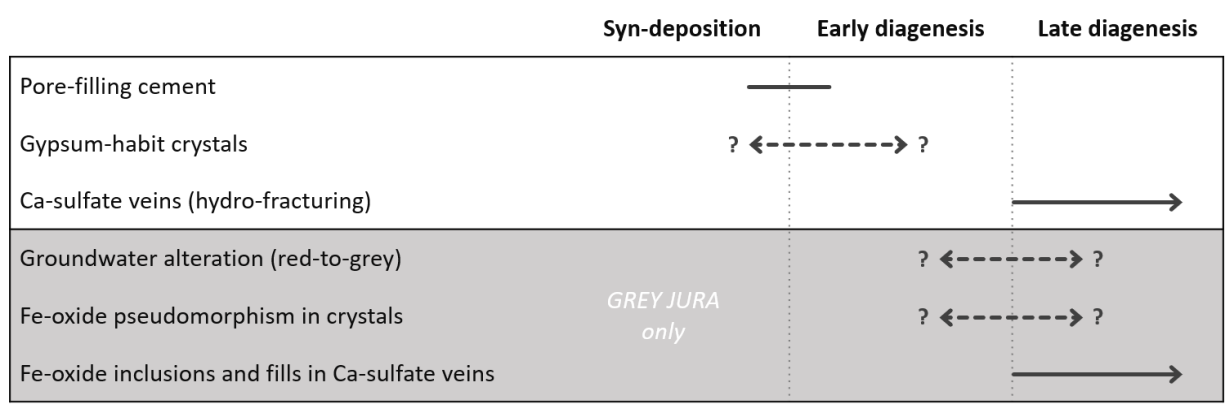

Figure 11: Paragenetic sequence of the events that occurred during the diagenesis of Vera Rubin ridge.

The VRR sedimentary rocks form a topographic high because they are harder and more resistant to erosion compared to the surrounding terrains. A possible explanation for such properties would be the presence of a fine-grained pore-filling cement, that formed early during the diagenesis of the VRR sedimentary rocks and resulted in their lithification and hardening. Moreover, the preservation of gypsum crystals in the red outcrops of the Jura member suggests that cementation was either contemporaneous with the formation of such crystals or involved limited fluid circulations in order to prevent the dissolution of the pre-existing evaporite minerals (Murray, 1964; Warren, 2016). As such, cementation likely occurred in the interstitial waters of the sediments in the early stages of diagenesis.

Lateral variations in outcrops observed on VRR (Figure 1), in the form of red to purple to grey color gradients, are attributed to a change in the mineralogy or crystallinity of these rocks (Horgan et al., this issue). Such lateral variations in outcrops do not appear to be related to changes in the depositional lacustrine environment (Edgar et al., this issue) and are thus inferred to be diagenetic in origin and involved groundwater interactions. Furthermore, the ability of fluids to propagate through the sedimentary rocks depends on its permeability, so the formation of the red-to-grey alteration patterns probably occurred before the complete compaction and lithification of the sediment reduced drastically its connected porosity (i.e. before late diagenesis). Lastly, the coarser grain sizes observed on the grey outcrops, compared to the adjacent finer-grained red outcrops (Rivera-Hernandez et al., 2019), would have also favored the propagation of fluids during diagenesis.

The timing of the formation of Ca-sulfate minerals may be difficult to assess as it can crystallize either by evapo-concentration near the surface or by reactions of dissolution and reprecipitation at depth due to its solubility, creating a variety of textures (Hardie, 1985). The gypsum-habit crystals observed within the VRR outcrops , displaying euhedral crystals morphologies (c.f. section 4.1.2), could either reflect unimpeded crystals growth before significant compaction and lithification of the host sediments, or instead formation of poikilotopic crystals (i.e., that incorporate sediment grains) through connected porosity in more consolidated sediments later-on during the diagenesis (Murray, 1964; Bain, 1990; Warren, 2016). Prior observations of Ca-sulfate enrichments in the Murray formation, below VRR in the stratigraphy, are interpreted to have formed near surface during early diagenesis due to the enrichment volume and lithology (Kah *et al.*, 2018; Rapin *et al.*, 2019). At VRR, the observed light-toned crystals represent a minor volume of the host rock and occur in a different lithology. Nevertheless, their formation must pre-date their substitution via pseudomorphism by dark-toned Fe-oxide within the grey outcrops (c.f. section 4.1.2). As such, the light-toned crystals were most likely already emplaced before the formation of red-to-grey color patterns within the VRR outcrops by groundwater interactions.

The light-toned Ca-sulfate veins are also observed throughout the VRR outcrops and are interpreted as late-diagenetic products formed by hydraulic fracturing (c.f. section 4.1.1), in line with previous observations along the rover traverse. However, the diversity of textures and morphologies observed in these features (Bennett *et al.*, this issue) could indicate multiple successive episodes of fracturing and mineralization (Caswell and Milliken, 2017; De Toffoli *et al.*, 2020), or distinct interactions with the enclosing host rocks.

Lastly, the dark-toned Fe-rich features encountered within the grey outcrops are often associated with light-toned Ca-sulfate veins, as either hexagonal inclusions (Figure 2 and 6) or fracture-fills (Figure 2 to 5), and appear to be contemporary with the formation of fractures and Ca-sulfate mineralization (c.f. section 4.1.3). Low-Fe bleached halos are also observed in association with the larger dark-toned Fe-rich features in the surrounding host rock (Figure 5), which suggests the mobilization of Fe at the time of the formation of dark-toned features. In addition, the Fe-rich dark-toned features that are not associated with light-toned veins display—for the most part—angular crystal shapes that are reminiscent of the gypsum-habit crystals observed in the red outcrops and are thus interpreted as pseudomorphs. The dark-toned pseudomorphs are not crosscut by light-toned veins (e.g. Figure 2C and Figure 4), which indicates that pseudomorphism likely occurred during or after the emplacement of light-toned veins. Since the substitution of light-toned crystals appears to be related to the formation of grey outcrops, the relative timing between fracture formation and groundwater interactions in the grey outcrops remains ambiguous, as the two events could be coeval.

In summary, the formation of the Vera Rubin ridge sedimentary deposits probably involved the following sequence of events (summarized in Figure 12):

1. deposition in a lacustrine setting;

2. cementation (c.f. section 5.2), either during sedimentation or within the interstitial pore-waters in the early stages of diagenesis (2A), which occurred either before or at the same time as the emplacement of gypsum euhedral crystals (2B);

3. formation of the red-to-grey color gradients within the outcrops of the ridge through groundwater alteration, which also led to the substitution of gypsum crystals by Fe-oxide pseudomorphs in the grey outcrops (3A), and likely contemporaneous with the formation of Ca-sulfate light-toned veins, associated in the grey outcrops with the mobilization of Fe from the adjacent host rock and the formation of Fe-oxide to hexagonal crystals and partial fills within the light-toned veins (3B).

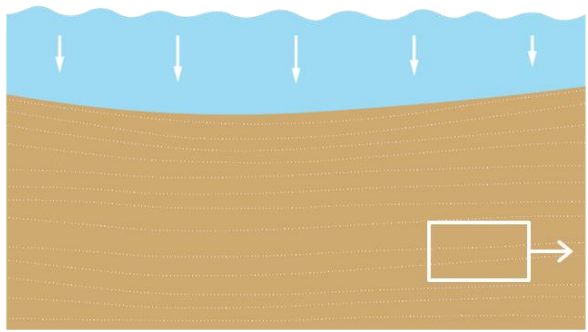
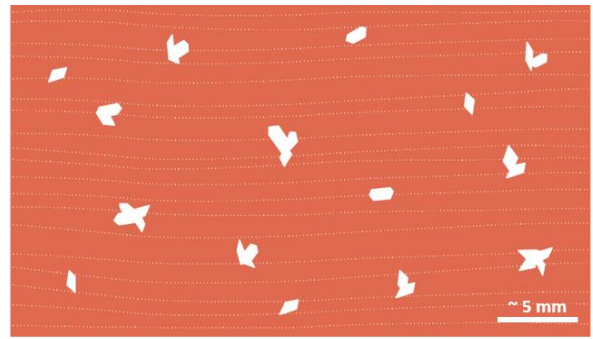

**1**    Deposition in lacustrine environment

**2A**    Cementation
**2B**    Formation of gypsum crystals

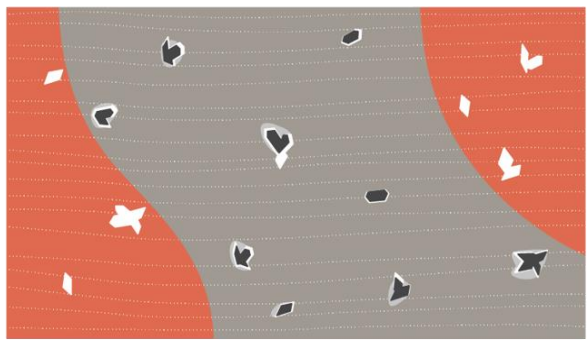
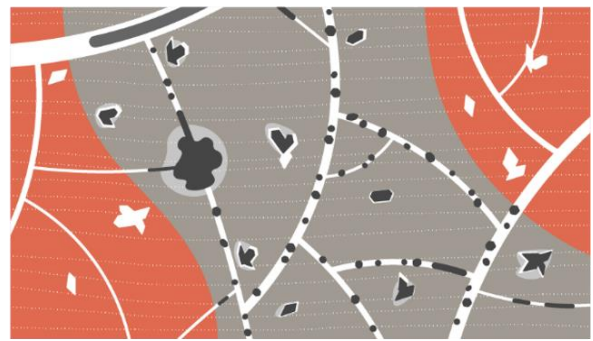

**3A**    Groundwater alteration in grey outcrops
Fe mobilization (bleaching) and formation of Fe-oxide pseudomorphs in crystal-shaped

**3B**    Fracturing and formation of light-toned veins
Fe mobilization (bleaching) and formation of Fe-oxide at the interface with Ca-sulfate veins

Figure 12: Schematic sequence of events involved in the formation of Vera Rubin ridge. Grey areas are represented schematically at the small scale for simplicity but can extend to several tens of meters. 3A and 3B are related to the mobilization of iron to form respectively dark-toned pseudomorphs and dark-toned inclusions in light-toned veins (euhedral crystals and partial fills), which may correspond to either a singular event or two distinct episodes, without any definitive evidence for the former or the latter.

## 5.2 Potential chemical processes during diagenesis

The formation of small-scale dark-toned Fe-oxide features on VRR is concentrated in the Jura member, suggesting that it experienced local diagenetic processes. In this section, we explore hypotheses that could explain the formation of the hematite through first (i) mobilization of Fe from the adjacent host rock leading to the formation of Fe-poor bleached halos, and second (ii) formation of Fe-oxide in association with Ca-sulfate diagenetic features, either as inclusions in light-toned veins or as mineral substitutes to pre-existing gypsum-habit crystals. While light-toned Ca-sulfate features are encountered in both red and grey outcrops throughout VRR, their association with dark-toned Fe-oxide features is almost exclusively restricted to the grey outcrops (c.f. section 4.1). As such, Fe mobilized from the adjacent grey host rock likely precipitated in the form of Fe-oxide upon interaction with Ca-SO$_4$ bearing fluids.

The most notable difference between red and grey outcrops—aside from their spectral properties and diagenetic components—lies in their respective Fe-bearing secondary mineralogy, with akaganeite and jarosite identified in association with red hematite in the red outcrops but not in the grey outcrops, which instead comprise grey crystalline hematite (Rampe et al., this issue). While the mobility of Fe observed in the grey outcrops could potentially result from the dissolution of metastable ferric mineral phases (i.e. jarosite, akaganeite) by groundwater fluids, their presence throughout the Jura member is contested as the drill location displayed peculiar chemical composition and spectral properties compared to other red outcrops (Frydenvang et al., this issue; David et al., this issue; Horgan et al., this issue). In addition, the dissolution of akageneite and jarosite alone – in the low abundances observed – could not account for the low-Fe abundances observed in the bleached halos (<10% wt.%). So, we conclude that other Fe-bearing minerals—notably hematite as the most abundant Fe-bearing mineral phase detected by CheMin in the VRR outcrops (Rampe et al., this issue)—had to be at least partially dissolved from the host rock to account for the observed mobility of Fe in the grey outcrops.

Generally speaking, hematite can be readily dissolved by acidic (low pH) or reducing (low Eh) conditions (Cornell and Schwertmann, 2003). In this regard, the detection of reduced sulfur in the drill samples from the Jura member by the SAM instrument would support the presence of reducing conditions in the diagenetic fluids (Wong et al., this issue, and references therein). The fact that both Fe and Mn have been mobilized in the grey outcrops, whereas non-redox sensitive elements (such as Si, Al, Mg, etc.) have not, also suggest the involvement of reducing fluids rather than strongly acidic ones. The observed variability in Fe and Mn across the VRR bedrocks (Frydenvang et al., this issue; David et al., this issue) have also been attributed to remobilization by reducing diagenetic fluids.

Reducing conditions are expected to have dominated the Noachian era of Mars, and to this day the Martian atmosphere does not contain high abundances of oxygen (e.g., Dehouck et al., 2016). Petrologic studies show that magmatic rocks on Mars are derived from a reducing mantle source, leading to the formation of reduced basalts (Herd et al., 2002). During sedimentation in the Gale crater, stratification in the lake waters could have also led to reducing conditions to prevail, at least locally (e.g., Hurowitz et al., 2017). In summary, it is fair to assume that reducing fluids could have carried Fe$^{2+}$ in solution during sedimentation (Figure 12 – Panel 1) and throughout the diagenesis (Figure 12 – Panel 2, 3A & 3B) at VRR. Moreover, reaction pathway models show that the assumption of 10 % Fe$^{2+}$ (over the total Fe) is sufficient to form hematite bearing, clay dominated assemblages upon diagenetic alteration of rocks with a bulk Murray-composition (e.g. Turner et al. this issue) (Figure 12 – Panel 2).

The partial dissolution of pre-existing hematite by reducing fluids during diagenesis would result in the release of Fe$^{2+}$ ions in the aqueous solution. In the absence of Ca-sulfate in close proximity, Fe$^{2+}$ mobilized in the host rock likely recrystallized in place as grey hematite, by coarsening of existing hematite grains through Ostwald ripening (David et al., this issue; Horgan et al., this issue) (Figure 12 – Panel 3A). This would explain the subdued ferric spectral properties and the mineralogy of the grey outcrops (Horgan et al., this issue; Rampe et al., this issue).

Conversely, $Fe^{2+}$ ions mobilized in fluids from the grey outcrops would be able to move toward the interface with Ca-SO$_4$ fluids following redox potential gradients (e.g., Chan et al., 2004, 2005). Indeed, as sulfate ions (SO$_4^{2-}$) present in the diagenetic fluids are an oxidizing agent (Barnes, 1997), the Ca-SO$_4$ bearing diagenetic fluids presented more oxidizing conditions compared to the surrounding host rock. Iron leached as $Fe^{2+}$ from the adjacent host rock oxidized to $Fe^{3+}$ and precipitated as grey crystalline hematite along with Ca-sulfate, in the form of substitution pseudomorphs to the gypsum-habit light-toned crystals (Figure 12 – Panel 3A) (e.g. Lougheed, 1983; Warren, 2016) and hexagonal crystals and partial fills in light-toned Ca-sulfate veins (Figure 12 – Panel 3B). In addition, the dissolution of hematite in the host rock could have been accelerated by the presence of SO$_4$-ions in the diagenetic fluids, as it would lead to an increased solubility of $Fe^{3+}$ through the formation of comparably stable $Fe^{3+}$-SO$_4$ complexes (Barnes, 1997).

The formation of hematite is also consistent with thermodynamic predictions, as illustrated by the Eh-pH diagram for the Fe-Ca-S system (Figure 13). Indeed, in this configuration, hematite is the most stable Fe-bearing phase over a broad range of pH and Eh conditions, except for reducing conditions where pyrite or pyrrhotite are expected to form instead (but neither are observed). In addition, laboratory experiments have shown that the presence of $Ca^{2+}$ ions in solution enhanced the crystallization of hematite (Mohapatra et al., 2012). In contrast, the formation of magnetite appears to be unlikely as it would require extremely alkaline and reducing fluid conditions.

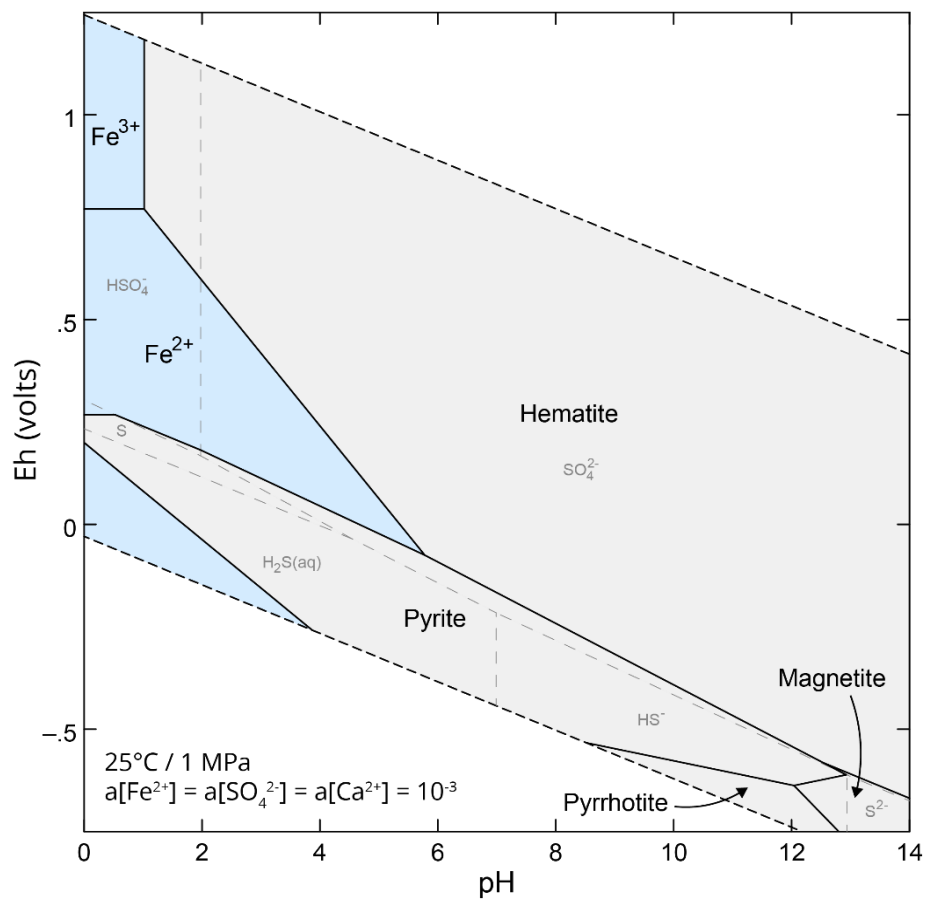

Figure 13: Eh-pH diagram for the Fe-S-Ca system, illustrating the thermodynamical stability fields of Fe-species (S species are indicated in light-grey). In this system, hematite constitutes the predominant phase except in reducing conditions, where pyrite is predicted to form, as well as pyrrhotite and magnetite in alkaline

conditions. The stability diagram was calculated using Geochemical Workbench and the LNLL thermodynamical database (Delany and Lundeen, 1990). The activity for all elements is fixed arbitrarily at $10^{-2}$.

## 5.3 Comparison with previous observations on Mars

The red-to-grey color gradients observed in the outcrops on Vera Rubin ridge bear similarities with the outcrops of the Burns sedimentary formation, explored by the *Opportunity* rover at Meridiani Planum. While it consists of evaporitic sandstones that formed in transient eolian and shallow water conditions (Squyres, 2004; Grotzinger *et al.*, 2005), the Burns formation outcrops also appear to have experienced an extended history of diagenetic alteration consistent with burial and periodic groundwater interaction(s) (McLennan *et al.*, 2005). There, syn-depositional and early-diagenetic intra-sedimentary evaporite crystal molds are observed, followed by at least two distinct episodes of cementation associated with interstitial pore-water and groundwater recharge(s) as well as the formation of hematite "blueberry" concretions (McLennan *et al.*, 2005; Squyres and Knoll, 2005). However, the diagenetic Fe-oxide features observed at Vera Rubin ridge display distinct morphologies with euhedral crystals, local pseudomorphs, and more complex polygonal shapes, compared to the well-rounded "blueberry" concretions observed at Meridiani, and as such originated from relatively different diagenetic processes. While hematite diagenetic features are only observed locally on VRR, and frequently in association with Ca-sulfate veins, the hematite concretions are pervasive throughout the outcrops at Meridiani Planum. One major difference resides in the fact that the sedimentary rocks of the Burns formation at Meridiani Planum incorporate significantly more sulfate (up to 40 wt.%), including Fe-sulfates such as jarosite and possibly melanterite (Squyres, 2004; Grotzinger *et al.*, 2005), which indicates that in this case, mobility of Fe during diagenesis was due to interaction with strongly acidic groundwater fluids (McLennan *et al.*, 2005; Tosca and McLennan, 2006). Although acidic fluids may have been involved in the formation of the VRR rocks on account of the jarosite observed at least locally at the Rock Hall drill hole, the predominance of Ca-sulfate over other sulfur-rich phases does not favor strongly acidic systems, and the detrital nature of the VRR sedimentary rocks (mudstones and fine-grained sandstones) make them significantly distinct from Meridiani Planum.

Earth analog studies have been conducted in order to provide insights into the diagenetic processes that occurred at Meridiani Planum, especially regarding the formation of hematite concretions. Among them, the extensive study of the Jurassic Navajo Sandstones (Chan, Parry and Bownman, 2000; Chan *et al.*, 2004, 2005, 2007; Ormö *et al.*, 2004; Beitler, Parry and Chan, 2005; Parry, Chan and Nash, 2009) is particularly relevant to Vera Rubin ridge. Indeed, outcrops there exhibit lateral color variations associated with the syn-sedimentary to early-diagenetic red hematite formation in oxidizing conditions and subsequent diagenetic alteration by reducing fluids, bleaching the outcrops white-grey at millimeter to regional scales (Chan *et al.*, 2004, 2005; Parry, Chan and Nash, 2009). Fe is either leached at that stage, incorporated into diagenetic fluids as $Fe^{2+}$ and later forming hematite concretions upon encountering oxidizing fluids (Chan *et al.*, 2005), or immediately fixed in ferroan dolomite, pyrite and large crystallites of grey hematite (Parry, Chan and Nash, 2009). Similar processes could also account for the lateral color variations observed on top of Vera Rubin ridge, with the early-diagenesis formation of ferric species (hematite, akaganeite, and jarosite) in oxidizing conditions and subsequent alteration by groundwater fluids, probably reducing, leading to the coarsening of hematite grains through dissolution/recrystallization processes and Ostwald ripening (David et al., this issue; Horgan et al., this issue, and references therein).

Thus, with Vera Rubin ridge and Meridiani Planum, the mobility of Fe during diagenesis has been observed in-situ at two locations on Mars, likely resulting from the interaction of groundwater fluids in two distinct settings, both involving hematite but also metastable Fe-bearing species such as Fe-sulfate (e.g. jarosite) and Fe-oxide/hydroxide (e.g. akaganeite). As such, orbital detections of akaganeite in evaporitic basins at Antoniadi and Robert Sharp (in addition to Gale) (Carter *et al.*, 2015; Buz *et al.*, 2017) and detections of jarosite in Valles Marineris (Milliken *et al.*, 2008), Noctis Labyrinthus (Thollot *et al.*, 2012), within the vicinity of Mawrth Vallis (Farrand *et al.*, 2009) and in the potential groundwater-fed paleolake deposits of Columbus crater at Terra Sirenum (Wray *et al.*, 2011) could constitute areas of interest for investigating groundwater interactions during diagenesis on Mars, involving the mobility of Fe. Orbital studies suggest that the diagenetic processes observed at Vera Rubin ridge may be relatively frequent on Mars, and our results show that the Meridiani Planum case should not be taken as the unique ground truth of such diagenetic systems.

## 6. Conclusion

The in-situ exploration of Vera Rubin ridge by the rover *Curiosity* has revealed a complex post-depositional geological history (Figure 12). The lateral color patterns observed in the VRR outcrops are attributed to syn-depositional or early diagenetic formation of ferric species (red color), and subsequent alteration by groundwater (to grey color). Within the grey outcrops, ChemCam highlighted that iron was locally mobilized from the host rock, likely by reducing groundwater fluids during diagenesis (as $Fe^{2+}$), to form ferric ($Fe^{3+}$) oxide as pseudomorphs of euhedral gypsum crystals, through coarsening of pre-existing Fe-oxides in the host rock, and as hexagonal crystalline inclusions and partial vein-fills within late-diagenetic Ca-sulfate fracture-fills. In the process, Fe was leached from the adjacent host rock upon interaction with $SO_4^{2-}$ bearing fluids, forming low-Fe bleached halos around the Fe-oxide diagenetic features. The formation of these diagenetic Fe-oxide minerals, interpreted as coarse-grained crystalline hematite, highlights the complex redox-driven processes that occurred during diagenesis and the induced mobility of redox-sensitive elements (mainly Fe, but also Mn to a lesser extent) occurring well after deposition and sedimentation. Thus, millimeter-scale observations from ChemCam show that diagenetic processes likely played a significant role in the formation of Vera Rubin ridge, a topographic structure visible from orbit.

**ACKNOWLEDGMENTS**

We are indebted to the Mars Science Laboratory Project engineering and science teams for their participations in tactical and strategic operations. All data used in this study are accessible at the Planetary Data Dystem: http://pds-geosciences.wustl.edu/missions/msl/index.htm. Data provided by the ChemCam instrument are supported in the US by NASA's Mars Exploration Program and in France by the Centre National d'Etudes Spatiales (CNES) and the Agence Nationale de la Recherche (ANR) under the program ANR-16-CE31-0012 entitled "Mars-Prime".

**SUPPLEMENTARY FIGURES**

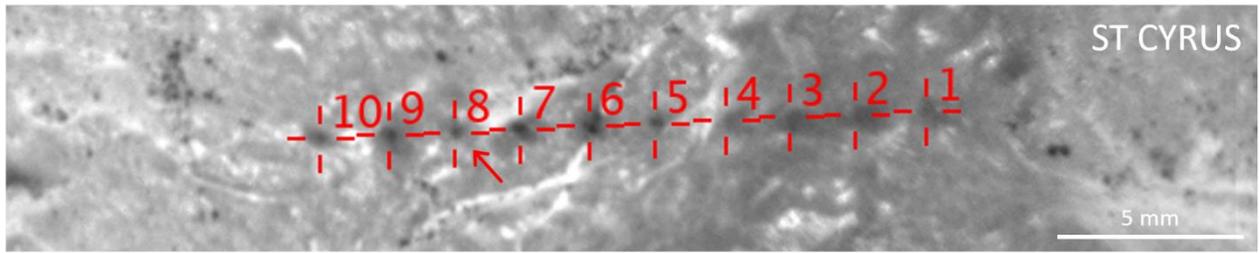

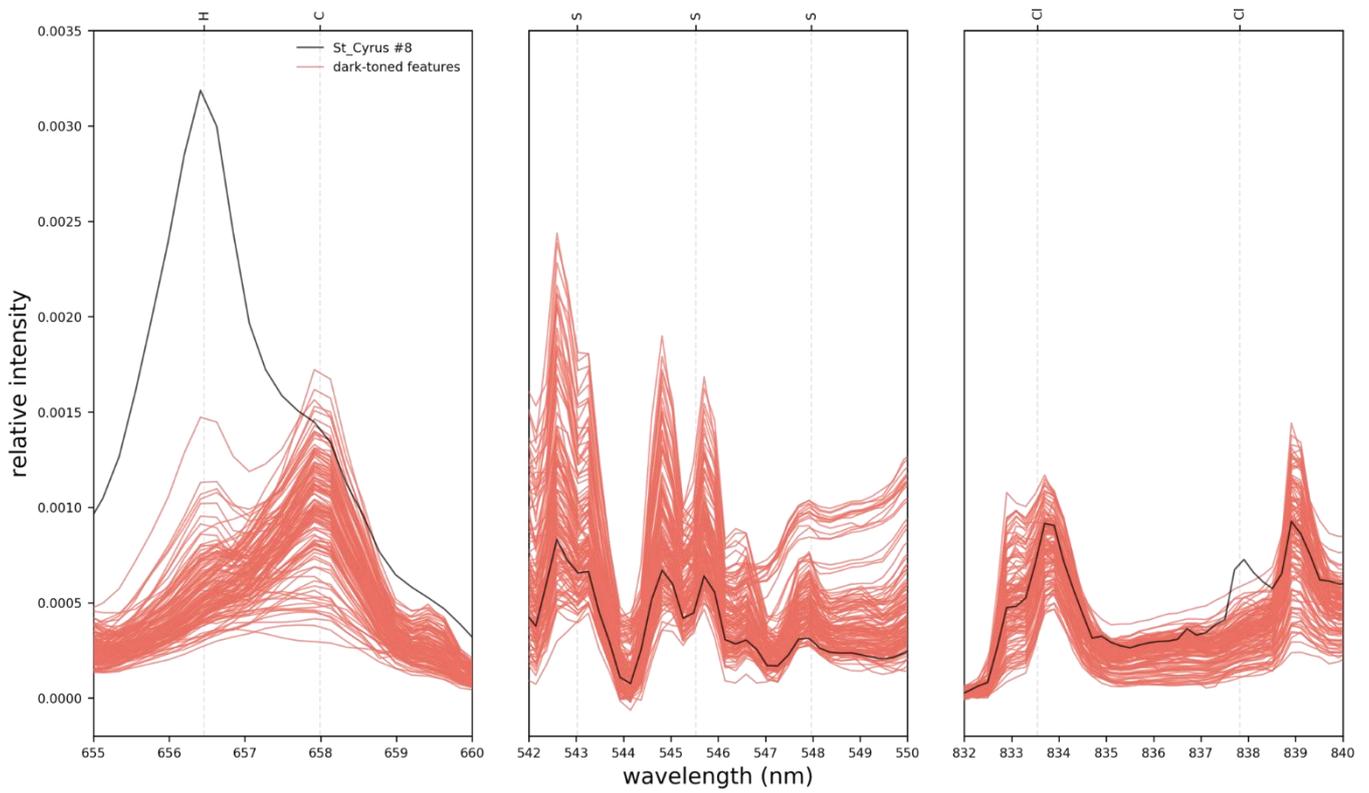

Figure S-1: LIBS spectra (normalized) of dark-toned features (red), showing the spectral region corresponding to the H, S, and Cl peaks location. St Cyrus #3 present high H and Cl peaks (at 837.9 nm), no S detection, as well as elevated Fe—thus suggesting the presence of akaganeite [$Fe^{3+}O(OH,Cl)$].

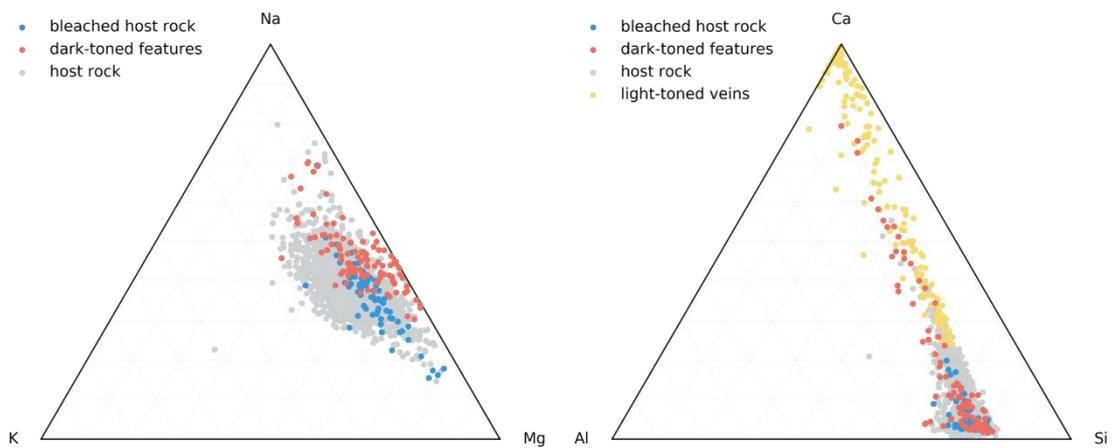

Figure S-2: Ternary diagrams showing ChemCam quantified abundances (molar percentages) for Na, K and Mg (left) as well as Ca, Al and Si (right). The trend toward the Ca summit is due to the sampling of Ca-sulfate veins, and the variability in K, Mg and Na is not correlated to the variability in Fe (either enrichments or depletions) as observed in Figure 5.